\documentclass[english,prl,reprint]{revtex4-2}
\usepackage[T1]{fontenc}
\usepackage[latin9]{inputenc}
\usepackage{color}
\usepackage{babel}
\usepackage{mathtools}
\usepackage{amsmath}
\usepackage{amssymb}
\usepackage{stmaryrd}
\usepackage{graphicx}
\usepackage[pdfusetitle,
 bookmarks=true,bookmarksnumbered=false,bookmarksopen=false,
 breaklinks=false,pdfborder={0 0 1},backref=false,colorlinks=true]
 {hyperref}
\hypersetup{
 pdfborderstyle=}

\makeatletter

\providecommand{\tabularnewline}{\\}

\usepackage{ragged2e}
\usepackage[justification=raggedright]{caption}
\DeclareCaptionJustification{justified}{\justifying}
\usepackage{colortbl}
\usepackage{tikz}
\usepackage{array}
\usepackage{tikz}
\usetikzlibrary{tikzmark}
\usetikzlibrary{decorations.pathreplacing,calligraphy}
\usepackage{orcidlink}

\usepackage{textcomp}
\usepackage{amsthm}
\newcommand*\LyXZeroWidthSpace{\hspace{0pt}}

\providecommand{\rrangle}{\rangle\!\rangle}
\usetikzlibrary{calc,arrows.meta}

\definecolor{macBlue}{RGB}{157,196,255}
\definecolor{macBlueText}{RGB}{70,110,180}
\definecolor{macPink}{RGB}{255,190,201}
\definecolor{macPinkText}{RGB}{190,90,110}
\definecolor{macMint}{RGB}{186,230,214}
\definecolor{macMintText}{RGB}{70,140,115}
\definecolor{macLavender}{RGB}{210,200,245}
\definecolor{macLavText}{RGB}{105,90,170}
\definecolor{macGray}{RGB}{245,245,245}
\definecolor{macBorder}{RGB}{160,160,160}
\definecolor{AccentBlue}{RGB}{82,110,160}
\definecolor{AccentBlueFill}{RGB}{244,247,252}
\definecolor{AccentRed}{RGB}{170,90,90}
\definecolor{AccentRedFill}{RGB}{252,247,247}
\definecolor{AccentGreen}{RGB}{85,125,100}
\definecolor{AccentGreenFill}{RGB}{247,251,248}
\definecolor{SoftGray}{RGB}{248,248,248}
\definecolor{MidGray}{RGB}{150,150,150}
\definecolor{DarkGray}{RGB}{70,70,70}

\ifdefined\showcaptionsetup
 \PassOptionsToPackage{caption=false}{subfig}
\fi
\usepackage{subfig}
\makeatother

\begin{document}
\title{Noise-Symmetry Optimization of Quantum Error-Corrected Metrology}
\author{Shuyun Su$^{1,2}$\orcidlink {0009-0000-5995-3277}}
\author{Shengshi Pang$^{1,3}$\orcidlink{0000-0002-6351-539X}}
\email{shengshp@mail.sysu.edu.cn}

\affiliation{$^{1}$School of Physics, Sun Yat-sen University, Guangzhou, Guangdong
510275, China\\$^{2}$State Key Laboratory for Mesoscopic Physics,
Department of Physics, Peking University, Beijing 100871, China\\$^{3}$Hefei
National Laboratory, University of Science and Technology of China,
Hefei 230088, China}
\begin{abstract}
Quantum error correction (QEC) code has emerged as a powerful tool
to protect quantum-enhanced metrology against noise. However, the
ability to correct errors alone does not guarantee high metrological
sensitivity, as the encoded states may become insensitive to the parameter
of interest. Here we show that this limitation can be overcome by
exploiting an intrinsic freedom of QEC codes: for a fixed set of correctable
errors, the Knill--Laflamme conditions admit an equivalence class
of encodings. When the correctable noise possesses unitary symmetries,
these symmetries generate continuous transformations within this class,
allowing systematic optimization of the encoding to increase the quantum
Fisher information while preserving the correctable set of noise.
Based on this observation, we develop a symmetry-based optimization
approach and derive criteria identifying when such optimization can
enhance metrological sensitivity. In particular, for stabilizer-sum
Hamiltonians, it shows that the symmetry optimization can convert
a code with vanishing QFI into one achieving the standard quantum
limit in general or even the Heisenberg scaling in specific cases,
illustrating the power of symmetry optimization for QEC-assisted quantum
metrology.
\end{abstract}
\maketitle
\global\long\def\hl{\mathcal{L}}%
\global\long\def\ha{\mathcal{A}}%
\global\long\def\hn{\mathcal{N}}%
\global\long\def\he{\mathcal{E}}%
\global\long\def\hr{\mathcal{R}}%
\global\long\def\hp{\mathcal{P}}%
\global\long\def\hj{\mathcal{J}}%
\emph{Introduction.}---Quantum metrology employs quantum effects,
such as quantum entanglement and squeezing, to enhance the precision
of parameter estimation beyond the standard quantum limit (SQL) in
noiseless systems \citep{np,paris2009quantum,BRAUNSTEIN1996135,PhysRevLett.96.010401,PhysRevLett.104.103602,Fujiwara2008AFB}.
These advantages have found broad applications ranging from gravitational
wave detection \citep{PhysRevA.87.032102,Schnabel2010QuantumMF,Abadie2011AGW},
quantum imaging \citep{PhysRevLett.110.050403,PhysRevA.79.013827,Brida2010ExperimentalRO,PhysRevLett.102.253601},
to quantum gyroscope \citep{PhysRevA.95.012326,PhysRevApplied.14.064023,PhysRevA.85.022333},
quantum target detection \citep{PhysRevResearch.3.L042039,Shapiro_2009,PhysRevLett.110.153603},
etc.

In practical applications, noise may spoil the quantum effects and
degrade measurement precisions. To address these challenges, conventional
approaches, e.g., adaptive measurements~\cite{Berry2000}, feedback
control~\cite{Fallani2022}, and stochastic measurement protocols~\cite{Muller2016},
suppress or mitigate errors during the sensing dynamics or readout
process. In contrast, quantum error correction (QEC) codes \citep{gottesman1997stabilizercodesquantumerror,Reiter2017,PhysRevResearch.4.023107,Lidar_Brun_2013},
which have been introduced to quantum metrology in recent years, exploit
the structure of the correctable noise to restore noiseless precision
and even recover the Heisenberg limit \citep{zhou2018achieving,vmd7-twd5,PhysRevA.106.052609,PhysRevLett.122.040502,PhysRevLett.112.080801}.
The QEC has been demonstrated for quantum metrology in experiments,
e.g., signal field sensing by a room-temperature hybrid spin \citep{unden2016quantum},
single-mode phase estimation in a superconducting circuit \citep{wang2019heisenberg},
frequency measurements on trapped ions \citep{roos2006designer},
magnetic field sensing by a 10-spin NOON state \citep{jones2009magnetic},
etc.

While QEC code is a powerful tool to protect quantum systems against
noise, the ability to correct errors does not by itself guarantee
good metrological performance \cite{PhysRevLett.133.190801}. A QEC
code that perfectly corrects the noise may remain weakly sensitive
or even completely insensitive to the parameter of interest, depending
on how the signal Hamiltonian acts within the code space. This reveals
that error correctability and metrological sensitivity are fundamentally
distinct requirements.

To address this issue, a key observation is that for a given set of
errors, quantum error correction does not uniquely determine the encoding.
In fact, the Knill-Laflamme theorem \citep{PhysRevA.55.900} allows
the existence of multiple QEC codes that correct the same set of noise,
which, nevertheless, may lead to drastically different precision for
quantum metrology. In particular, if the set of correctable errors
is invariant under specific unitary transformations, these transformations
can generate a family of QEC codes that have the same error-correcting
capability but different metrological sensitivity. Such unitary symmetries
therefore provide the possibility to enhance the metrological sensitivity
of QEC codes.

Inspired by the above idea, we develop a symmetry-based approach for
optimizing the metrological sensitivity of QEC codes in this Letter.
We characterize the symmetries of correctable noise and provide a
constructive method to find their generators, which are then applied
to maximize the quantum Fisher information (QFI) for an unknown parameter
encoded in the Hamiltonian. We further derive criteria that determine
when such optimization is possible. We demonstrate the power of this
approach in stabilizer-sum Hamiltonians, where the logical code space
coincides with the Hamiltonian ground space and yields zero QFI but
symmetry-based optimization can turn such metrologically \textquotedblleft dark\textquotedblright{}
codes into highly sensitive ones without sacrificing the error-correcting
ability. Importantly, we clarify how the achievable precision scales
with the number of physical qubits, a question that has remained largely
unexplored in QEC-assisted metrology, and provide a general lower
bound showing that symmetry-optimized stabilizer codes can attain
at least SQL scaling, a seemingly modest scaling but actually nontrivial
and not guaranteed a priori. Moreover, we show that more favorable
scalings, including Heisenberg scaling, can be attained when only
a finite number of symmetry transformations is required.

\emph{Preliminaries.}---In the theory of parameter estimation, Fisher
information determines the fundamental precision limit of estimating
an unknown parameter. For a parameter $\text{\ensuremath{\theta}}$
encoded in a probability distribution $P(X|\theta)$ of a random variable
$X$, the Fisher information is defined as the second moment of the
score function $\partial_{\theta}\log P(X|\theta)$ \citep{fisher_1925,doiSailes1995},
and the precision of any unbiased estimator of $\theta$ is limited
through the Cram\'er-Rao bound \citep{rao1992information,BRAUNSTEIN1996135,Phy,taylor1979cramer}
by
\begin{equation}
\Delta^{2}\hat{\theta}\cdot F(\theta)\geq1,
\end{equation}
where $\Delta^{2}$ denotes the variance.

In quantum mechanics, the parameter $\theta$ is usually encoded in
a quantum state and can be inferred by measurements on the system.
The Fisher information of the parameter depends on the measurement
schemes, and maximizing the Fisher information over all possible measurement
schemes yields the quantum Fisher information, which characterizes
the ultimate estimation precision allowed by quantum mechanics \citep{BRAUNSTEIN1996135,Care_1983,Phy}.
For a pure state $\rho_{\theta}=|\psi_{\theta}\rangle\langle\psi_{\theta}|$
generated by unitary evolution under a Hamiltonian $\theta H_{0}$
for an evolution time $t$, $|\psi_{\theta}\rangle=e^{-i\theta H_{0}t}|\psi\rangle$,
where $|\psi\rangle$ is the initial state of the system, the QFI
can be obtained as
\begin{equation}
\mathcal{F}_{\theta}=4t^{2}\left\langle \psi\left|\Delta^{2}H_{0}\right|\psi\right\rangle .\label{eq:fisher}
\end{equation}
And it can be shown that $\mathcal{F}_{\theta}$ is maximized to $t^{2}(\lambda_{\max}-\lambda_{\min})^{2}$
when $|\psi{}_{0}\rangle$ is an equal superposition of the eigenstates
of $H_{0}$ with the maximum and minimum eigenvalues $\lambda_{\max}$
and $\lambda_{\min}$.

We now briefly recall the theory of QEC code. The basic idea of QEC
is encoding logical information into a subspace of the quantum system
such that different errors map the code space to orthogonal subspaces.
Let $\he$ be a set of error operators $\{\varepsilon_{a}\}$ correctable
by an $\llbracket n,l,d\rrbracket$ QEC code encoding $l$ logical
qubits into $n$ physical qubits with code distance $d$. The correctability
of $\he$ requires the existence of a trace-preserving quantum channel
$\hr$ that reverses $\he$ for any state $\rho$ in the code space,
i.e., $(\hr\circ\he)(\rho)\propto\rho$, where the proportionality
factor is independent of $\rho$. 
\begin{figure}
\includegraphics[width=0.6\columnwidth]{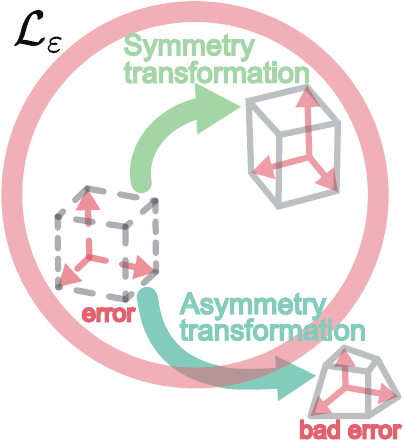}

\caption{Symmetry of correctable error space. A symmetry transformation (light
green arrow) keeps the correctable errors within the linear space
spanned by themselves (red circle). In contrast, an asymmetry transformation
(deep green arrow) drives a correctable error out of $\mathcal{L}_{\varepsilon}$
and turns it into an uncorrectable error.}
\end{figure}
 This leads to the Knill-Laflamme (KL) theorem \citep{PhysRevA.55.900,PhysRevA.52.R2493,PhysRevLett.77.793}
which gives the necessary and sufficient condition for the existence
of $\hr$ that corrects $\he$,
\begin{equation}
P\varepsilon^{\dagger}_{a}\varepsilon_{b}P=C_{ab}P,\label{eq:qecc}
\end{equation}
with $P$ the projector onto the code space and $C_{ab}$ the elements
of a Hermitian matrix $C$.

\emph{Continuous symmetry of noise}.---Symmetry plays a fundamental
role in physics \citep{laborde2021testing,faist2020continuous,mintun2015bulk}.
If a system is invariant under a group of unitary transformations
$U(q)=\text{exp[-i}Aq]$, where $q$ is a real parameter and $A$
is a Hermitian operator, it can be regarded to possess the symmetry
generated by $A$. For a continuous symmetry group, any infinitesimal
transformation $U(\delta q)\approx\mathbb{I}-\text{i}A\delta q$ keeps
the state of the system invariant.

Now we apply the continuous symmetry to QEC. Consider a set of errors
$\{\varepsilon_{a}\}$, which is correctable by a QEC code with $P$
as the projector onto the code space. $P$ and $\{\varepsilon_{a}\}$
satisfy the KL theorem \eqref{eq:qecc}. When the noise is transformed
by a general unitary operator $U$, the KL theorem is also satisfied
by a transformed code $UPU^{\dagger}$,
\begin{equation}
UPU^{\dagger}(U\varepsilon_{a}U^{\dagger})^{\dagger}(U\varepsilon_{b}U^{\dagger})UPU^{\dagger}=C_{ab}UPU^{\dagger}.
\end{equation}
Due to the discretization of errors in QEC, any combination of the
correctable errors $\{\varepsilon_{a}\}$ can also be corrected by
the same code, implying all the correctable errors span a linear space,
which we denote as $\hl_{\varepsilon}$. Now, if the correctable error
space $\hl_{\varepsilon}$ is invariant under $U$ (though each $U\varepsilon_{a}U^{\dagger}$
can be different from $\varepsilon_{a}$), i.e., $U^{\dagger}\mathcal{\hl}_{\varepsilon}U=\mathcal{\hl}_{\varepsilon}$,
the error set $\{\varepsilon_{a}\}$ is then also correctable by the
transformed QEC code with the code space $UPU^{\dagger}$, which means
a new code is generated by the symmetry of noise that can correct
the same noise!

To exploit the continuous symmetry of noise to optimize the QEC code
for quantum metrology, it is crucial to find all the continuous symmetries
possessed by the correctable noise. It can be shown that for any continuous
symmetry possessed by the error space $\hl_{\varepsilon}$, a generator
$A$ of the symmetry must satisfy
\begin{equation}
[A,\mathcal{\hl}_{\varepsilon}]\subseteq\mathcal{\hl}_{\varepsilon},\,A=A^{\dagger}.\label{eq:ada}
\end{equation}
The proof is given in Sec. I of the Supplemental Material~\cite{SM}.

In this work, we focus on the independent noise model in which errors
act independently on different physical qubits and the noise rate
is sufficiently low. In this case, it can be shown that the symmetry
generators of the correctable noise space $\hl_{\varepsilon}$ are
all one-body local operators, i.e., spanned by single-qubit operators,
and the correctability of the QEC code remains invariant under any
local unitary transformation. The proof is provided in Sec. II of
the Supplemental Material~\cite{SM}.

\emph{QEC code optimization for QFI}.---The existence of noise symmetries
implies that a family of QEC codes can correct the same set of errors
but exhibit distinctive metrological performance. This freedom allows
optimization of the QEC code to enhance the sensitivity of quantum
measurements without sacrificing the error correcting ability.

We first illustrate this idea with the three-qubit bit-flip code as
an example. Consider a Hamiltonian $H(\theta)=-\theta(Z_{1}Z_{2}+Z_{2}Z_{3}+Z_{3}Z_{1})$,
with an unknown parameter $\theta$ to estimate. The conventional
choice of code space is spanned by $\{|000\rangle,|111\rangle\}$,
which is actually a degenerate ground space of the Hamiltonian, so
the variance of the Hamiltonian within the code space vanishes and
the QFI for $\theta$ is zero, rendering the code metrologically inactive.
Nevertheless, the bit-flip noise subspace is invariant under arbitrary
single-qubit unitary transformations, so we can rotate the code space
by a unitary operation, e.g., $\exp(\text{i}\pi/4X_{2})$, to $\{(|000\rangle+\text{i}|010\rangle)/\sqrt{2},(|111\rangle+\text{i}|101\rangle)/\sqrt{2}\}$.
This rotated code can  correct exactly the same  bit-flip errors as
the original one, but the encoded states are no longer eigenstates
of the Hamiltonian and thus acquire nonzero variances of $H(\theta)$.
In fact, each of the two basis states can now lead to the highest
QFI $16t^{2}$ after an evolution of time $t$ according to Eq. \eqref{eq:fisher},
demonstrating that noise symmetry can activate a metrologically \textquotedblleft dark\textquotedblright{}
QEC code and restore Heisenberg scaling!

Motivated by this observation, we now formulate the general QEC code
optimization problem for quantum metrology. Let $\hn(\hl_{\varepsilon})$
denote the symmetry generator space of the error space $\hl_{\varepsilon}$
with an orthonormal basis $\{F_{a}\}$. The symmetry transformations
are generated by $\sum_{i}q_{i}F_{i}$ with $\{q_{i}\}$ real coefficients,
which define a family of QEC codes with the same error correctability.
To reach the highest precision for the parameter $\theta$, one needs
to optimize the coefficients $\{q_{i}\}$ to maximize the QFI.

In the following, we focus on two fundamental questions regarding
the optimization of QEC codes: (i) for what kind of Hamiltonians the
QEC codes can be optimized to increase the QFI; (ii) the scaling of
QFI when the QEC codes are optimized. We will give general criteria
for Hamiltonians that allow effective optimization of QEC codes and
show that the scaling of QFI with the number of physical qubits can
be significantly increased.

\emph{Hamiltonian criteria for symmetry optimization.}---For QEC
codes, we find criteria determining whether a code can be effectively
optimized for quantum metrology to increase the QFI via noise symmetry
by a variational approach. Specifically, for a Hamiltonian $H(\theta)=\theta H_{0}$,
the general first-order condition that allows effective optimization
of a QEC code can be derived as
\begin{equation}
\Pi_{\mathcal{N}(\mathcal{L}_{\epsilon})}\left[H_{0},\rho\Delta H_{0}+\Delta H_{0}\rho\right]\neq0,\label{eq:HSO}
\end{equation}
where $\rho$ is the density matrix of the code state, $\Delta H_{0}=H_{0}-\langle H_{0}\rangle$,
and $\Pi_{\mathcal{N}(\mathcal{L}_{\epsilon})}$ is the projector
onto the admissible symmetry generator space $\mathcal{N}(\mathcal{L}_{\epsilon})$.
We refer to this condition as the first-order Hamiltonian symmetry
optimization (HSO) condition. For independent Pauli errors, this condition
can be simplified to that the commutator $\left[H_{0},\rho\Delta H_{0}+\Delta H_{0}\rho\right]$
contains at least one single-body operator. When the first-order variation
vanishes, a second-order optimization condition can be obtained. Detailed
derivation and results are provided in Sec. III of the Supplemental
Material~\cite{SM}.

We now consider stabilizer codes as a typical example. Stabilizer
codes are one of the most widely-used classes of QEC codes based on
stabilizer groups. A stabilizer group is an Abelian subgroup of a
multi-qubit Pauli group, and the code space of a stabilizer code is
the simultaneous +1 eigenspace of all stabilizers. For a stabilizer-sum
Hamiltonian \citep{PhysRevLett.123.230503},
\begin{equation}
H(\theta)=\theta H_{0},\;H_{0}=-\sum_{i}g_{i},\label{eq:smHami}
\end{equation}
where $\{g_{i}\}$ are the generators of the underlying stabilizer
group $\mathcal{S}=\left\langle g_{1},\cdots,g_{n-l}\right\rangle $
that defines the code and $\theta$ is the unknown parameter to estimate,
the code space coincides with the ground space of $H(\theta)$, therefore,
the logical code states remain invariant under $H(\theta)$, yielding
vanishing QFI for $\theta$. 
\begin{figure}
\begin{centering}
\subfloat[\label{fig:QFI-with-rotation}QFI with symmetry transformation of
code space. \centering]{\includegraphics[width=4.2cm]{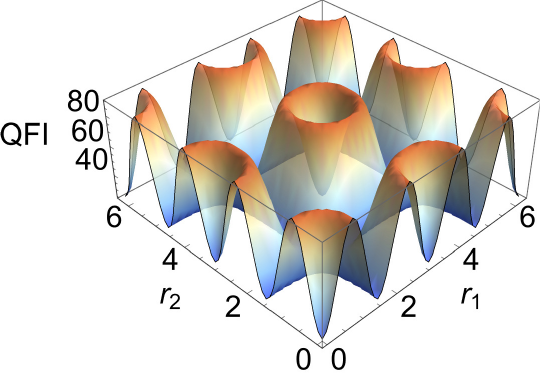}

}\subfloat[\label{fig:Optimization-of-QFIs}Optimization of QFI with respect
to $\vec{r}_{1}(q)$ and $\vec{r}_{2}(q)$.\centering]{\begin{centering}
\includegraphics[width=0.5\columnwidth]{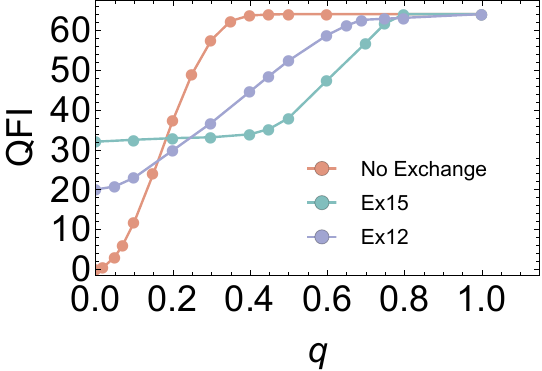}
\par\end{centering}
}
\par\end{centering}
\subfloat[Optimal control paths of $\vec{r}_{1}(q)$ with and without qubit
exchange.]{\includegraphics[height=3.8cm]{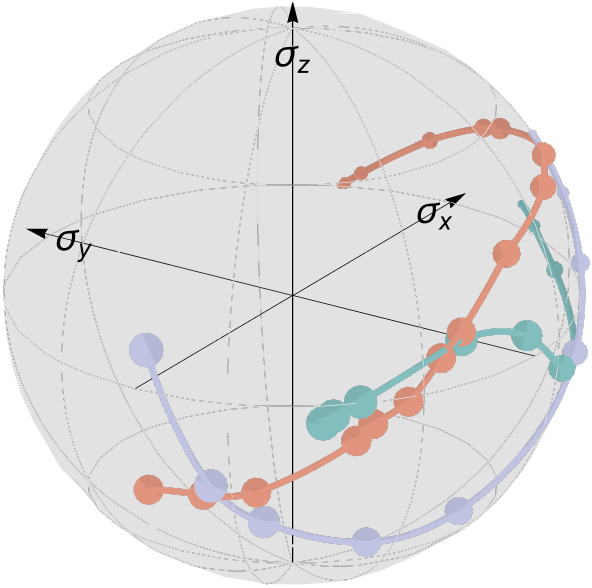}

}\subfloat[Pauli components of optimal control path $\vec{r}_{1}(q)$ without
qubit exchange \centering]{\includegraphics[height=3.2cm]{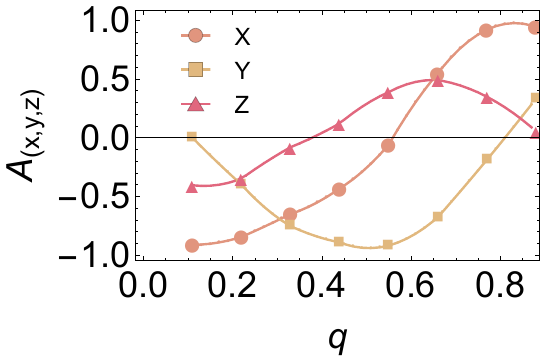}

}

\caption{Symmetry optimization of five-qubit code for stabilizer-sum Hamiltonian.
(a) QFI is initially zero, but can rise to the maximum via local symmetry
transformation $U(q)=\bigotimes_{k}e^{\text{i}\vec{r}_{k}(q)\cdot\vec{\sigma}}$,
where $\vec{r}_{1}(q)$ and $\vec{r}_{2}(q)$ correspond to the first
and second qubits, respectively. (b) The efficacy of continuous symmetry
optimization can be enhanced by qubit exchange operations, such as
exchanging the first and fifth qubits (labeled as \textquotedblleft$\text{Ex}15$\textquotedblright )
or exchanging the first and second qubits (labeled as \textquotedblleft$\text{Ex}12$\textquotedblright ).
(c) The optimal control paths $\vec{r}_{1}(q)$ are plotted for different
scenarios, and share the legend with (b). The radius of each point
on the control paths represents the estimation error. (d) The Pauli
components of optimal control path $\vec{r}_{1}(q)$ are plotted without
qubit exchange.}\label{fig:Symmetry-optimization-of}
\end{figure}
Nevertheless, it can be verified that $H(\theta)$ generally meets
the second-order HSO criterion given in Sec. III of the Supplemental
Material~\cite{SM} under independent errors and hence the noise symmetry
can rotate the associated stabilizer code into a new, equivalent one
whose logical states are no longer eigenstates of $H(\theta)$. As
a result, the encoded states acquire nonzero variance of the Hamiltonian
and become sensitive to the parameter $\theta$, while preserving
the same error-correction capability.
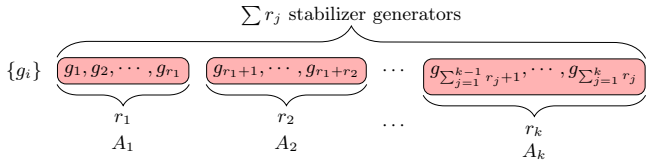
\begin{figure}
\raggedright \scalebox{0.8}{ \begin{tikzpicture}

% Define styles
\tikzset{
    box/.style={rectangle, draw, fill=red!30, rounded corners},
    brace/.style={decorate, decoration={brace, amplitude=10pt}},
    label/.style={font=\small}
}

% Draw boxes
\node[box] (A1) at (0, 0) {${g_1, g_2, \cdots, g_{r_1}}$};
\node[box] (A2) at (2.7, 0) {${g_{r_1+1}, \cdots, g_{r_1+r_2}}$};
\node[box] (Ak) at (6.8, -0.1) {${g_{\sum_{j=1}^{k-1} r_j + 1}, \cdots, g_{\sum_{j=1}^k r_j}}$};

% Draw braces
\draw[brace]    (A1.south east)-- (A1.south west) node[midway, left=5pt, label] {$ $};
\draw[brace]      (A2.south east)--  (A2.south west) node[midway, left=5pt, label] {$ $};
\draw[brace]      (Ak.south east)  --(Ak.south west) node[midway, left=5pt, label] {$ $};

% Labels under boxes
\node at (0, -1.2) {$A_1$};
\node at (2.7, -1.2) {$A_2$};
\node at (6.8, -1.33) {$A_k$};

\node at (0, -0.8) {$r_1$};
\node at (2.7, -0.8) {$r_2$};
\node at (6.8, -0.99) {$r_k$};

\node at (4.5, 0) {$\cdots$};
\node at (4.5, -0.9) {$\cdots$};

% Braces at the top
\draw [decorate,decoration={brace,mirror,amplitude=15pt},yshift=15pt]  (Ak.north east)--(A1.north west)  node [midway, yshift=20pt] {$\sum r_j$ stabilizer generators};

	\node at (-1.64, 0) {$\{g_i\}$};
%	\node at (0, 0) {000};

\end{tikzpicture}}

\caption{Each $A_{j}$ anticommutes with $r_{j}$ stabilizer generators only.
Together, the $k$ symmetry generators $A_{j}$ anticommute with $\sum^{k}_{j=1}r_{j}$
stabilizer generators jointly.}\label{fig:generatorof control}
\end{figure}

As an explicit example, the five-qubit QEC code $\llbracket5,1,3\rrbracket$,
the smallest code that can protect against any single-qubit error
with the stabilizers given in Table \ref{tab:five-qubit-code2}, yields
zero QFI for the stabilizer-sum Hamiltonian \eqref{eq:smHami}. Nevertheless,
a local symmetry generator $Y_{4}$ anticommutes with all the stabilizer
generators, so it can turn the ground space into the eigenspace associated
with the largest eigenvalue of all the stabilizer generators. Consequently,
$\text{exp}(\text{i}\pi/4Y_{4})$ turns a ground state to an equal
superposition of the lowest and highest eigenstates, producing the
maximum variance of $H(\theta)$ and hence the maximum QFI, as shown
in Fig. \ref{fig:Symmetry-optimization-of} and detailed in Sec. VII
of the Supplemental Material~\cite{SM}.

\emph{QFI scaling.}---Previous studies have shown that the QEC codes
can restore Heisenberg-limited precisions with the number of \emph{logical}
qubits \citep{PhysRevLett.112.080801}. For large-scale QEC codes,
the scaling of measurement precision with the number of \emph{physical}
qubits becomes a critical and nontrivial question. In the following,
we show that the symmetry optimization of stabilizer QEC codes guarantees
at least SQL scaling with the number of physical qubits, and achieves
the Heisenberg scaling in favorable scenarios.

Consider a stabilizer-sum Hamiltonian associated with a stabilizer
QEC code $\llbracket n,l,d\rrbracket$. While the symmetry transformation
is generally generated by the linear combination of the symmetry generators,
it is instructive to take a step-by-step approach to optimize the
QEC codes, which will clearly show how the optimization proceeds.

We first consider the ideal case that there exists a Hermitian operator
$A$ satisfying the noise symmetry condition \eqref{eq:ada} and anticommuting
with all stabilizers simultaneously. In this case, $A$ can flip the
eigenvalues of all stabilizers and map a ground state of the Hamiltonian
to the highest excited state. So, the unitary transformation $\exp(-\text{i}\frac{\pi}{4}A)$
converts an initial code state into an equal superposition of the
lowest and highest levels of the Hamiltonian and attains the highest
estimation precision for the parameter $\theta$ of the Hamiltonian,
with the error-correction capability invariant.

This idea can be further generalized to the scenario where several
symmetry generators contribute to the optimization. Suppose there
are $k$ mutually commuting Hermitian operators $\{A_{j}\}$ satisfying
the noise symmetry condition and each $A_{j}$ anticommutes with a
distinct subset of the stabilizer generators. If these subsets are
disjoint and collectively cover all the stabilizer generators, one
can apply each $\exp(-\text{i}q_{j}A_{j})$ to the initial state,
where $q_{j}$ is a real number, and derive the composite symmetry
transformation as
\begin{equation}
U=\prod^{k}_{j=1}\text{exp}(-\text{i}q_{j}A_{j}).
\end{equation}
Let each $A_{j}$ anticommute with $r_{j}$ stabilizer generators
of the QEC code, as shown in Fig. \ref{fig:generatorof control}.
Then the stabilizers are partitioned into disjoint subsets with $\sum^{k}_{j=1}r_{j}=n-l\propto O(n)$.
By applying the symmetry optimization to a ground state of the stabilizer-sum
Hamiltonian \eqref{eq:smHami}, it can be shown that the QFI is lower
bounded by
\begin{equation}
(\mathcal{F}_{\theta})_{\max}\ge4t^{2}\frac{(n-l)^{2}}{k}.\label{eq:fmax}
\end{equation}
Therefore, when $k$ remains constant as the system size increases,
the estimation precision reaches the Heisenberg limit with respect
to the number of physical qubits $n$! Detail of the derivation is
provided in Sec. IV of the Supplemental Material~\cite{SM}.

This construction is particularly suitable for the stabilizer codes
of which all the stabilizer generators are composed of Pauli $X$
and $Z$ operators only. In this case, one can first choose $A_{1}=Y_{i_{1}}$,
acting only on the $i_{1}$-th qubit that has the most $X$ and $Z$
operators across all the stabilizer generators and thus $A_{1}$ anticommutes
with the generators with $X$ or $Z$ in that qubit. Then one can
choose $A_{2}=Y_{i_{2}}$ , acting only on the $i_{2}$-th qubit that
has the most $X$ and $Z$ operators in the remaining generators.
One can iterate this process, and finally divide all the stabilizers
into disjoint subsets anticommuting with different $A_{j}$'s and
all $A_{j}$'s commute with each other.

We illustrate this process by some typical stabilizer codes. For the
five-qubit code, the stabilizer generators of which are shown in Table
\ref{tab:five-qubit-code2}, a single-qubit operator $Y_{4}$ suffices
to anticommute with all the stabilizer generators. For the nine-qubit
Shor code encoding one logical qubit, whose stabilizer generators
are shown in Table \ref{tab:The-stabilizer-generators-of-nine-qubit-code},
it requires at least 3 operators $\{X_{2,}Y_{5},X_{8}\}$ to anticommute
with all the stabilizer generators.

As the number $k$ of the symmetry generators cannot exceed the number
$n-l$ of the stabilizer generators, Eq. \eqref{eq:fmax} guarantees
at least SQL scaling $O(n)$ of QFI. Whenever a finite number of symmetry
generators, independent of $n$, collectively flip all stabilizer
generators, the QFI can further reach Heisenberg scaling $O(n^{2})$.
\begin{table}
\subfloat[Stabilizer generators of five-qubit code\label{tab:five-qubit-code2}]{%
\begin{tabular}{c|ccccc}
\hline 
\multicolumn{1}{c||}{$g_{i}$} & \multicolumn{5}{c}{Operator}\tabularnewline
\hline 
$g_{1}$ & $X$ & $Z$ & $Z$ & $\textcolor{red}{X}$ & $I$\tabularnewline
$g_{2}$ & $I$ & $X$ & $Z$ & $\textcolor{red}{Z}$ & $X$\tabularnewline
$g_{3}$ & $X$ & $I$ & $X$ & $\textcolor{red}{Z}$ & $Z$\tabularnewline
$g_{4}$ & $Z$ & $X$ & $I$ & $\textcolor{red}{X}$ & $Z$\tabularnewline
\hline 
\end{tabular}

}\qquad{}\subfloat[Stabilizer generators of nine-qubit code\label{tab:The-stabilizer-generators-of-nine-qubit-code}.]{%
\begin{tabular}{c|ccccccccc}
\hline 
$g_{i}$ & \multicolumn{9}{c}{Operator}\tabularnewline
\hline 
$g_{1}$ & $Z$ & $\textcolor{red}{Z}$ & $I$ & $I$ & $I$ & $I$ & $I$ & $I$ & $I$\tabularnewline
$g_{2}$ & $I$ & $\textcolor{red}{Z}$ & $Z$ & $I$ & $I$ & $I$ & $I$ & $I$ & $I$\tabularnewline
$g_{3}$ & $I$ & $I$ & $I$ & $Z$ & $\textcolor{red}{Z}$ & $I$ & $I$ & $I$ & $I$\tabularnewline
$g_{4}$ & $I$ & $I$ & $I$ & $I$ & $\textcolor{red}{Z}$ & $Z$ & $I$ & $I$ & $I$\tabularnewline
$g_{5}$ & $I$ & $I$ & $I$ & $I$ & $I$ & $I$ & $Z$ & $\textcolor{red}{Z}$ & $I$\tabularnewline
$g_{6}$ & $I$ & $I$ & $I$ & $I$ & $I$ & $I$ & $I$ & $\textcolor{red}{Z}$ & $Z$\tabularnewline
$g_{7}$ & $X$ & $X$ & $X$ & $X$ & $\textcolor{red}{X}$ & $X$ & $I$ & $I$ & $I$\tabularnewline
$g_{8}$ & $I$ & $I$ & $I$ & $X$ & $\textcolor{red}{X}$ & $X$ & $X$ & $X$ & $X$\tabularnewline
\hline 
\end{tabular}

}\caption{Stabilizer generators of typical QEC codes for symmetry-based metrological
optimization. (a) The five-qubit $\llbracket5,1,3\rrbracket$ code.
A single local noise symmetry generator $Y_{4}$ anticommutes with
all stabilizer generators. (b) The nine-qubit Shor code. A single
local operator is insufficient to anticommute with all stabilizers.
Instead, a set of three mutually commuting operators, $\{X_{2},Y_{5},X_{8}\}$,
is required to collectively anticommute with the entire set of stabilizer
generators. In both panels, the Pauli operators of the stabilizer
generators that anticommute with the chosen local symmetry generators
are highlighted in red, which identify the specific physical qubits
where the symmetry transformation acts to rotate the metrologically
dark code space into a highly sensitive state and maximizes the QFI
while exactly preserving the error-correction capability of the code.}

\label{tab:1}
\end{table}
In more general cases, intermediate scalings are also possible. Sec.
VII of the Supplemental Material~\cite{SM} presents an example with
$\mathcal{F}_{\theta}=O(n^{\frac{3}{2}})$. Thus, the achievable scaling
is governed by how the number of required symmetry generators grows
with the system size, which is of particular significance for large-scale
QEC codes, e.g., the surface code \citep{Krinner2022,Sundaresan2023,Acharya2025,zhaoRealizationErrorCorrectingSurface2022a},
that have received growing attention in recent years.

In practice, symmetry transformations may be incorporated into the
encoding circuit prior to sensing. When they are implemented by physical
controls, their time cost, calibration errors, and induced noise should
be included in the metrological resource count \cite{PhysRevLett.128.140503},
making the achievable precision enhancement platform- and implementation-dependent.
Meanwhile, although the present work is focused on independent noise,
the approach is fundamentally formulated in terms of the correctable
error space and can accommodate correlated, crosstalk, or higher-order
errors by enlarging this space and identifying its corresponding symmetries
\cite{wfyl-wtz3}. The resulting optimization then depends on the
symmetry structure of the enlarged error space, but the symmetry optimization
principle remains unchanged.

\emph{Conclusions}.---In this Letter, we show that noise symmetry
provides a useful resource for optimizing QEC--assisted quantum metrology.
Although a QEC code may perfectly correct a given noise set, it can
remain metrologically inactive. By characterizing the continuous symmetries
of the correctable errors, we demonstrate that these symmetries generate
families of QEC codes that are equivalent for error correction but
inequivalent for metrology. Based on this observation, we develop
a symmetry optimization framework and derive a criterion that determines
when metrological sensitivity can be activated. For stabilizer codes,
this approach turns metrologically dark code states into sensitive
ones, guarantees at least SQL scaling of QFI with respect to the number
of physical qubits and enables enhanced scalings, including the Heisenberg
scaling, in favorable scenarios.

\emph{Acknowledgments.}--- This work is supported by the National
Natural Science Foundation of China (Grant No. 12075323), the Natural
Science Foundation of Guangdong Province of China (Grant No. 2025A1515011440)
and the Innovation Program for Quantum Science and Technology (Grant
no. 2021ZD0300702).

\begingroup
\makeatletter
\let\addcontentsline\@gobblethree
\makeatother
\endgroup
\makeatother
\makeatletter
% Both reference lists are embedded above/below.  Prevent REVTeX from looking
% for an additional jobname.bbl at \end{document}.
\global\let\auto@bib\@empty
\makeatother

% ============================================================================
% Supplemental Material (single-source arXiv layout)
% ============================================================================
\clearpage
\onecolumngrid
\setcounter{page}{1}
\renewcommand{\thepage}{S\arabic{page}}
\setcounter{section}{0}
\setcounter{subsection}{0}
\setcounter{equation}{0}
\setcounter{figure}{0}
\setcounter{table}{0}
\renewcommand{\thesection}{\Roman{section}}
\renewcommand{\thesubsection}{\Alph{subsection}}
\renewcommand{\theequation}{S\arabic{equation}}
\renewcommand{\thefigure}{S\arabic{figure}}
\renewcommand{\thetable}{S\arabic{table}}
\renewcommand{\theHequation}{S.\arabic{equation}}
\renewcommand{\theHfigure}{S.\arabic{figure}}
\renewcommand{\theHtable}{S.\arabic{table}}
\renewcommand{\theHsection}{S.\arabic{section}}

\global\long\def\ho{o}%
\global\long\def\hu{\mathcal{U}}%
\global\long\def\hs{\mathcal{S}}%
\global\long\def\i{{\rm i}}%
\global\long\def\hf{\mathcal{F}}%

\begin{center}
{\large\bfseries Supplemental Material for \textquotedblleft Noise-Symmetry
Optimization of Quantum Error-Corrected Metrology\textquotedblright\par}
\vspace{0.8em}
Shuyun Su$^{1,2}$\orcidlink{0009-0000-5995-3277}\quad
Shengshi Pang$^{1,3}$\orcidlink{0000-0002-6351-539X}\par
\vspace{0.5em}
{\small
$^{1}$School of Physics, Sun Yat-sen University, Guangzhou, Guangdong 510275, China\\
$^{2}$State Key Laboratory for Mesoscopic Physics, Department of Physics,
Peking University, Beijing 100871, China\\
$^{3}$Hefei National Laboratory, University of Science and Technology of China,
Hefei 230088, China\\
shengshp@mail.sysu.edu.cn
}
\end{center}
\vspace{0.8em}

\tableofcontents
\vspace{1em}

Quantum metrology exploits quantum resources to enhance the sensitivity
of measurements and exceed the classical precision limit. However,
noise can severely degrade these quantum advantages. Quantum error
correction (QEC) has been developed to effectively protect quantum
systems against noise and preserve useful information. This work studies
the continuous symmetry of noise and applies it in QEC codes to optimize
the code space and maximize the quantum Fisher information (QFI) for
estimation of parameters in Hamiltonians. In this Supplemental Material,
we provide detailed derivations for the main results in the main text.
We first derive the continuous noise symmetry condition, and characterize
the structure of symmetry generators. We then develop a general symmetry-based
optimization framework and establish the criteria of Hamiltonian symmetry
optimization (HSO) to increase QFI by a variational approach. We further
derive the achievable QFI bounds under symmetry optimization and analyze
their scalings with the number of physical qubits. Finally, we illustrate
the principle of symmetry optimization through explicit constructions
and typical examples of stabilizer codes, including detailed analyses
of the surface code, the two-dimensional cruciform code, and the five-qubit
code, etc.

\section{Generators of continuous symmetry\label{SM:sec:Generators-of-continuous}}

In this section, we derive the generators for the continuous symmetry
of noise. Consider a QEC code $\llbracket n,l,d\rrbracket$ and let
$\mathcal{R}$ be the correction superoperator that can correct the
noise channel $\mathcal{E}$, i.e., $\mathcal{R\circ\mathcal{\mathcal{E}}}(P)=P$,
where $P$ is the projector onto the code space. If the correctable
noise space $\mathcal{L_{\varepsilon}}$ is invariant under a unitary
transformation $U$, i.e.,
\begin{equation}
U^{\dagger}\mathcal{L}_{\varepsilon}U=\mathcal{L_{\varepsilon}},\label{SM:eq:s2}
\end{equation}
where $\mathcal{L_{\varepsilon}}$ is the linear space spanned by
the correctable noise $\{\varepsilon_{a}\}$.

For a clear representation and efficient computation of superoperators,
we employ the bivector formalism. For example, the bivector form of
$\varepsilon_{a}$ is denoted as $\left|\varepsilon_{a}\right\rrangle $
which is the vector by stacking the columns of $\varepsilon_{a}$
from left to right. The bivector form of Eq. \eqref{SM:eq:s2} is
\begin{equation}
U^{\dagger}\otimes U^{\intercal}\left|\varepsilon_{a}\right\rrangle \in V_{\varepsilon},\quad\forall\left|\varepsilon_{a}\right\rrangle \in V_{\varepsilon},
\end{equation}
where the superscript $\intercal$ denotes matrix transposition and
$V_{\varepsilon}$ is the subspace spanned by bivectors $\{\left|\varepsilon_{1}\right\rrangle ,\left|\varepsilon_{2}\right\rrangle ,\left|\varepsilon_{3}\right\rrangle \cdots\}$.
This equation can also be rewritten in a more compact way as
\begin{equation}
(\mathbb{I}-Q)(U^{\dagger}\otimes U^{\intercal})Q=\mathbf{0},
\end{equation}
where $Q$ is the projection operator onto the bivector subspace $V_{\varepsilon}$,
which gives
\begin{equation}
Q(U^{\dagger}\otimes U^{\intercal})Q=(U^{\dagger}\otimes U^{\intercal})Q.\label{SM:eq:5}
\end{equation}
Taking the Hermitian conjugate on both sides of this equation, we
have
\begin{equation}
Q(U^{\dagger}\otimes U^{\intercal})^{\dagger}Q=Q(U^{\dagger}\otimes U^{\intercal})^{\dagger}.\label{SM:eq:5-1}
\end{equation}
Multiplying the above two equations leads to 
\begin{equation}
Q\Pi Q=Q,
\end{equation}
where $\Pi\coloneqq(U^{\dagger}\otimes U^{\intercal})Q(U^{\dagger}\otimes U^{\intercal})^{\dagger}$
is the unitarily transformed projection operator. On the other hand,
since the unitary operator $U^{\dagger}\otimes U^{\intercal}$ does
not change the dimension of $Q$, it is evident that
\begin{equation}
Q\Pi Q\le Q.\label{SM:eq:8}
\end{equation}
The equality in Eq. \eqref{SM:eq:8} holds if and only if $\Pi=Q$, hence
it turns out that
\begin{equation}
[Q,U^{\dagger}\otimes U^{\intercal}]=\mathbf{0}.\label{SM:eq:pup}
\end{equation}
So, $Q$ and $U^{\dagger}\otimes U^{\intercal}$ share the same eigenbasis.

Suppose $U=e^{\text{i}qA}$, where $A$ is a Hermitian matrix with
eigenvalues $\lambda_{k}$ and eigenstates $|\lambda_{k}\rangle$
and $q$ is a real parameter. The eigendecomposition of $U^{\dagger}\otimes U^{\intercal}$
is
\begin{equation}
U^{\dagger}\otimes U^{\intercal}=\sum_{a,b}e^{\text{i}q(\lambda_{b}-\lambda_{a})}\left|\lambda_{a},\lambda^{*}_{b}\right\rangle \left\langle \lambda_{a},\lambda^{*}_{b}\right|.
\end{equation}
Using this decomposition, Eq. \eqref{SM:eq:pup} can be simplified to
\begin{equation}
[Q,\text{exp}(-\text{i}q\mathcal{M}[\mathbf{ad}_{A}]\text{)}]=\mathbf{0},\label{SM:eq:pep}
\end{equation}
where $\mathbf{\mathcal{M}}[\mathbf{ad}_{A}]$ denotes the matrix
representation of the superoperator $\mathbf{ad}_{A}(\cdot)\coloneqq[A,\cdot]$,
given by
\begin{equation}
\mathcal{M}[\mathbf{ad}_{A}]=\sum_{a,b}(\lambda_{a}-\lambda_{b})\left|\lambda_{a},\lambda^{*}_{b}\right\rangle \left\langle \lambda_{a},\lambda^{*}_{b}\right|.
\end{equation}
For continuous symmetry, $q$ can be any real value. In this case,
it can be inferred from Eq. \eqref{SM:eq:pep} that
\begin{equation}
[Q,\mathcal{M}[\mathbf{ad}_{A}]]=\mathbf{0},
\end{equation}
which is equivalent to
\begin{equation}
\mathbf{ad}_{A}(\varepsilon_{a})=[A,\varepsilon_{a}]\in\mathcal{L}_{\varepsilon},\forall\varepsilon_{a}\in\mathcal{L}_{\varepsilon},\label{SM:eq:ada}
\end{equation}
or in a more compact form,
\begin{equation}
[A,\mathcal{\hl}_{\varepsilon}]\subseteq\mathcal{\hl}_{\varepsilon}.\label{SM:eq:ada-1}
\end{equation}

It implies that $\mathcal{L}_{\varepsilon}$ is an invariant subspace
of $\mathbf{ad}_{A}$, i.e., $\mathcal{L}_{\varepsilon}\in\text{Inv}(\mathbf{ad}_{A})$.
Eq. \eqref{SM:eq:ada-1} provides the condition for the generators $A$
of continuous symmetry given the set of correctable errors. It can
be seen that all instances of $A$ satisfying Eq. \eqref{SM:eq:ada-1}
form a real vector space, due to the Hermiticity of symmetry generators,
which we denote as $\mathcal{N}(\mathcal{L_{\varepsilon}})$ throughout
this paper, i.e., 
\begin{equation}
\hn(\mathcal{L}_{\epsilon}):=\{A=A^{\dagger}\,|\,[A,\mathcal{L}_{\epsilon}]\subseteq\mathcal{L}_{\epsilon}\}.
\end{equation}

\section{Derivation of $\mathcal{N}(\mathcal{L_{\varepsilon}})$}

In Sec. \ref{SM:sec:Generators-of-continuous}, we derive the condition
for the symmetry generators of general noise models. In this section,
we focus on the independent noise model and the QEC code $\llbracket n,l,d\rrbracket$
correcting the errors of weight at most $w_{\max}=\left\lfloor \frac{d-1}{2}\right\rfloor $.
The correctable error space $\mathcal{L_{\varepsilon}}$ is hence
spanned by all Pauli strings of weight at most $w_{\max}$.

We now analyze the structure of the symmetry-generator space $\mathcal{N}(\mathcal{L}_{\varepsilon})$.
In the following, we consider two distinct scenarios depending on
whether the symmetry generators $\{A_{j}\}$ in $\mathcal{N}(\mathcal{L_{\varepsilon}})$
are one-body local operators, i.e., those spanned by single-qubit
operators, or multi-body operators. It turns out that Eq. \eqref{SM:eq:ada}
is satisfied only by the former case. Thus, all symmetry generators
$\{A_{j}\}$ must be one-body local operators.

Scenario I.--- We assume that the symmetry generators $\{A_{j}\}$
in $\mathcal{N}(\mathcal{L_{\varepsilon}})$ are all one-body local
operators. Equivalently, any symmetry generator $A\in\mathcal{N}(\mathcal{L}_{\varepsilon})$
can be expressed as a real linear combination of operators acting
nontrivially on individual qubits. Therefore, without loss of generality,
we may choose a basis $\{A_{j}\}$ of $\mathcal{N}(\mathcal{L}_{\varepsilon})$
such that each basis element $A_{j}$ acts nontrivially only on a
single qubit. Thus, the QEC code is invariant under the symmetry group
generated by local transformations
\begin{equation}
U=\bigotimes^{n}_{j=1}U_{j},
\end{equation}
where $U_{j}=e^{\text{i}A_{j}}$. It is evident that all single-qubit
$\{A_{j}\}$ satisfy Eq. \eqref{SM:eq:ada} and the local unitary transformation
$U$ does not increase the weight of any $n$-qubit Pauli strings.
So, the errors after the symmetry transformations are still inside
$\hl_{\varepsilon}$, i.e., $U\varepsilon_{i}U^{\dagger}\in\hl_{\varepsilon}$.
This confirms that all single-qubit operators belong to $\mathcal{N}(\mathcal{L}_{\varepsilon})$
and generate continuous symmetries of the correctable noise space.

Scenario II.--- We next assume by contradiction that there exists
a symmetry generator $A$ in $\mathcal{N}(\mathcal{L_{\varepsilon}})$
that contains at least one nontrivial multi-qubit Pauli string. $A$
can be generally expanded in the Pauli basis,
\begin{equation}
A=\sum_{P}a_{P}P,\label{SM:eq:Aexp}
\end{equation}
where the summation runs over all $n$-qubit Pauli strings. Since
different Pauli strings may have support on different qubits, the
condition $\left[A,\hl_{\varepsilon}\right]\subseteq\hl_{\varepsilon}$
needs to be analyzed component by component. Moreover, as $A$ could
contain more than one Pauli string, we also need to account for the
possibility of cancellations among contributions generated by different
commutators. In particular, terms lying outside $\mathcal{L}_{\varepsilon}$
may cancel in the summation even if they appear in individual Pauli-string
contributions.

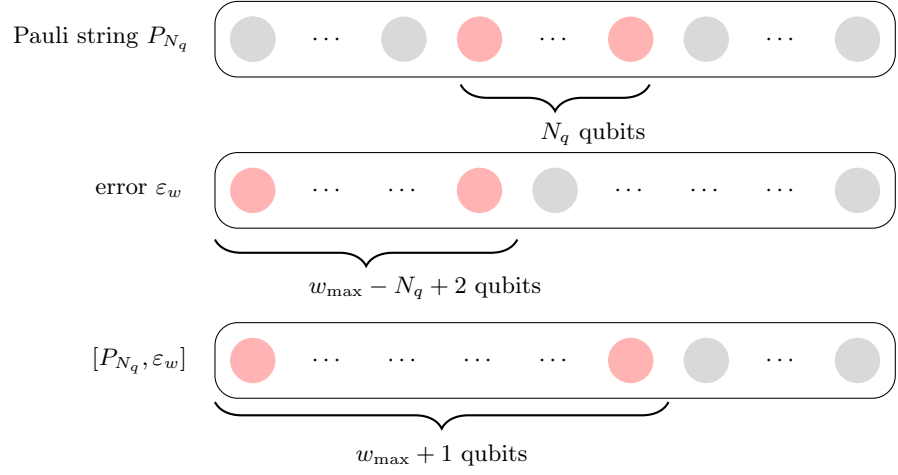
\begin{figure}[h]
\begin{tikzpicture}
\tikzset{
    box/.style={rectangle, draw, rounded corners},
    brace/.style={decorate, decoration={brace, amplitude=10pt}},
    mylabel/.style={font=\small}
}

\draw[rounded corners=3mm] (-0.5,2-0.5) rectangle (8.5,2.5);
\draw[rounded corners=3mm] (-0.5,-0.5) rectangle (8.5,0.5);

\draw[thick,decorate,decoration={brace,amplitude=10pt}] (3.5,-0.65)--(-0.5,-0.65);
\node at (2.28, -1.29) {$w_{\max}-N_{q}+2$ qubits};

% A_j
\draw[thick,decorate,decoration={brace,amplitude=10pt}] (5.25,2-0.6)--(2.75,2-0.6);
\node at (4.5, 2-1.25) {$N_q$ qubits};
\fill[gray!30] (0,2) circle (0.3);
\node at (1, 2) {$\cdots$};
\fill[gray!30] (2,2) circle (0.3);
\fill[gray!30] (0,2) circle (0.3);
\fill[red!30] (3,2) circle (0.3);
\node at (4, 2) {$\cdots$};
\fill[red!30] (5,2) circle (0.3);
\fill[gray!30] (6,2) circle (0.3);
\node at (7, 2) {$\cdots$};

% E'
\fill[red!30] (0,0) circle (0.3);
\node at (1, 0) {$\cdots$};

\node at (2, 0) {$\cdots$};
\fill[red!30] (3,0) circle (0.3);
\fill[gray!30] (4,0) circle (0.3);
\node at (5, 0) {$\cdots$};
\node at (6, 0) {$\cdots$};

\node at (7, 0) {$\cdots$};
\fill[gray!30] (8,0) circle (0.3);

\fill[gray!30] (8,2) circle (0.3);

% [A_j, E'_e]
\fill[red!30] (0,-2.25) circle (0.3);

\node at (1, -2.25) {$\cdots$};
\node at (2, -2.25) {$\cdots$};
\node at (3, -2.25) {$\cdots$};

\fill[red!30] (5,-2.25) circle (0.3);
\fill[gray!30] (6,-2.25) circle (0.3);
\node at (4, -2.25) {$\cdots$};
\node at (7, -2.25) {$\cdots$};
\fill[gray!30] (8,-2.25) circle (0.3);

\draw[rounded corners=3mm] (-0.5,-2.75) rectangle (8.5,-1.75);
\draw[thick,decorate,decoration={brace,amplitude=10pt}] (5.5,-2.8)--(-0.5,-2.8);
\node at (2.5, -3.5) {$w_{\max}+1$ qubits};

% left labels
\node at (-2, 2.0) {Pauli string $P_{N_{q}}$};
\node at (-1.5, 0) {error $\varepsilon_{w}$};
\node at (-1.5, -2.25) {$[P_{N_q},\varepsilon_{w}]$};

\end{tikzpicture}

\caption{A Pauli string $P_{N_{q}}$ of weight $N_{q}\protect\geq2$ anticommutes
with an error operator $\varepsilon_{w}$ of weight $w_{\max}-N_{q}+2$
on at least one qubit in the support of $\varepsilon_{w}$. Thus,
$[P_{N_{q}},\varepsilon_{w}]$ is supported on $w_{\max}+1$ qubits.\label{SM:fig:The-control-}}
\end{figure}

Suppose $\mathcal{L}_{\varepsilon}$ contains all Pauli strings up
to the weight $w_{\max}$ and $A$ contains a Pauli component $P_{N_{q}}$
of weight $N_{q}\geq2$. If $N_{q}>w_{\max}$, one can choose a single-qubit
error $\varepsilon_{1}$ that anticommutes with a qubit in the support
of $P_{N_{q}}$, $\mathrm{supp}(P_{N_{q}})$, i.e., the set of physical
qubits on which $P_{N_{q}}$ acts nontrivially. In this case, $[P_{N_{q}},\varepsilon_{1}]=2P_{N_{q}}\varepsilon_{1}$
is still of weight $P_{N_{q}}>w_{\max}$, and therefore it does not
belong to $\hl_{\varepsilon}$. On the other hand, if $2\leq N_{q}\leq w_{\max}$,
one can then find a correctable Pauli error $\varepsilon_{w}$ of
weight $w=w_{\max}-N_{q}+2\leq w_{\max}$ so that it anticommutes
with $P_{N_{q}}$ on one of the qubits in $\mathrm{supp}(P_{N_{q}})$
and has its remaining support of $w_{\max}-N_{q}+1$ qubits outside
$\mathrm{supp}(P_{N_{q}})$. As illustrated in Fig. \ref{SM:fig:The-control-},
the commutator $[P_{N_{q}},\varepsilon_{w}]$ introduces additional
nontrivial Pauli operators from $\mathrm{supp}(P_{N_{q}})$ on the
qubits where $\varepsilon_{w}$ originally acted trivially. As a result,
the commutator $[P_{N_{q}},\varepsilon_{w}]=2P_{N_{q}}\varepsilon_{w}$
is a Pauli string of weight $w_{\max}+1$, therefore, this commutator
lies outside $\hl_{\varepsilon}$, i.e.,
\begin{equation}
[P_{N_{q}},\varepsilon_{w}]\notin\mathcal{L_{\varepsilon}}.
\end{equation}

In either case, the map $P\rightarrow P\varepsilon$, $\varepsilon=\varepsilon_{1}\,{\rm or}\,\varepsilon_{w}$,
is a bijection on $P$, so the Pauli string $2P_{N_{q}}\varepsilon$
can be induced only from the $P_{N_{q}}$ itself and cannot be generated
by any other Pauli component in the expansion \eqref{SM:eq:Aexp} of
$A$. Consequently, the contribution $\left[P_{N_{q}},\varepsilon\right]=2P_{N_{q}}\varepsilon$
cannot cancel with any other term in $[A,\varepsilon]$. Hence, $[A,\varepsilon]$
necessarily contains a component of weight at least $w_{\max}+1$.
This contradiction rules out the existence of any multi-qubit symmetry
generator in $\mathcal{N}(\mathcal{L}_{\varepsilon})$.

Combining the above two scenarios, we prove that the symmetry generator
space $\hn(\hl_{\varepsilon})$ for independent noise is spanned by
single-qubit operators. Consequently, for independent noise at sufficiently
low error rates, the correctable noise space $\mathcal{L}_{\varepsilon}$
is invariant under the continuous symmetry group generated by the
local transformations $U=\bigotimes^{n}_{j=1}U_{j}$, where the unitary
operator $U_{j}$ acts only on the $j$-th qubit.

\section{Hamiltonian symmetry optimization criteria}

In the main text, the Hamiltonian symmetry optimization (HSO) is introduced
as the enhancement of metrological sensitivity achieved by exploiting
continuous symmetries of the correctable error space. In this section,
we derive the HSO criterion to increase the sensitivities of QEC codes
in detail.

\subsection{First-order HSO criterion}

Suppose the code space is $\mathcal{C}$, and let $|\psi\rangle\in\mathcal{C}$
be the initial code state. The parameter is encoded through the Hamiltonian
$H(\theta)=\theta H_{0}$. For a pure state under unitary parameter
encoding, the QFI with respect to the parameter $\theta$ is
\begin{equation}
\hf_{\theta}(\psi)=4t^{2}{\rm Var}_{\psi}(H_{0}),
\end{equation}
where ${\rm Var}_{\psi}(X)=\langle\psi|X^{2}|\psi\rangle-\langle\psi|X|\psi\rangle^{2}$.
For a symmetry transformation
\begin{equation}
U(\boldsymbol{q})=\exp\left(-\i\sum^{M}_{\mu=1}q_{\mu}A_{\mu}\right),\;A_{\mu}\in\mathcal{N}(\mathcal{L}_{\epsilon}),
\end{equation}
where $\{A_{\mu}\}$ is an orthonormal basis of the admissible symmetry
generator space $\mathcal{N}(\mathcal{L}_{\epsilon})$, the QFI after
the symmetry rotation of the code is
\begin{equation}
\hf_{\theta}(\boldsymbol{q},\psi)=4t^{2}{\rm Var}_{U(\boldsymbol{q})\psi}(H_{0}).
\end{equation}
For the convenience of analysis, we write this QFI in an equivalent
way,
\begin{equation}
\hf_{\theta}(\boldsymbol{q},\psi)=4t^{2}{\rm Var}_{\psi}K(\boldsymbol{q}),
\end{equation}
where $K(\boldsymbol{q})=U(\boldsymbol{q})^{\dagger}H_{0}U(\boldsymbol{q}).$
The symmetry optimization problem is therefore to determine whether
there exists $\boldsymbol{q}$ such that
\begin{equation}
\hf_{\theta}(\boldsymbol{q},\psi)>\hf_{\theta}(\boldsymbol{0},\psi).
\end{equation}
Obviously, establishing such a general symmetry optimization condition
would require a global maximization of the QFI $\mathcal{F}_{\theta}$
over all possible symmetry transformations, which is generally difficult.
Here, we resort to the perturbative analysis and derive the first-
and second-order conditions for the symmetry optimization by a variational
approach. Higher-order conditions can be derived in a similar way.

Let $\rho_{\psi}=|\psi\rangle\langle\psi|$, $\Delta H_{0}=H_{0}-\langle H_{0}\rangle_{\psi}I$.
For any single symmetry generator $A_{\mu}\in\mathcal{N}(\mathcal{L}_{\epsilon})$,
define $K_{A_{\mu}}(q)=e^{\i qA_{\mu}}H_{0}e^{-\i qA_{\mu}}$. The
first derivative of the variance is
\begin{equation}
\left.\frac{\partial}{\partial q_{\mu}}{\rm Var}_{\psi}[K_{A_{\mu}}(q)]\right|_{\boldsymbol{q}=0}=2\i{\rm Cov}_{\psi}\left(H_{0},[A_{\mu},H_{0}]\right),
\end{equation}
where ${\rm Cov}_{\psi}(X,Y)=\frac{1}{2}\langle XY+YX\rangle_{\psi}-\langle X\rangle_{\psi}\langle Y\rangle_{\psi}$.
Using cyclicity of the trace, this derivative can be written as
\begin{equation}
2\i{\rm Cov}_{\psi}\left(H_{0},[A_{\mu},H_{0}]\right)={\rm Tr}\left[A_{\mu}G_{\psi}(H_{0})\right],
\end{equation}
with $G_{\psi}(H_{0})=\i\left[H_{0},\rho_{\psi}\Delta H_{0}+\Delta H_{0}\rho_{\psi}\right]$,
where the imaginary unit $\i$ makes $G_{\psi}(H_{0})$ Hermitian.

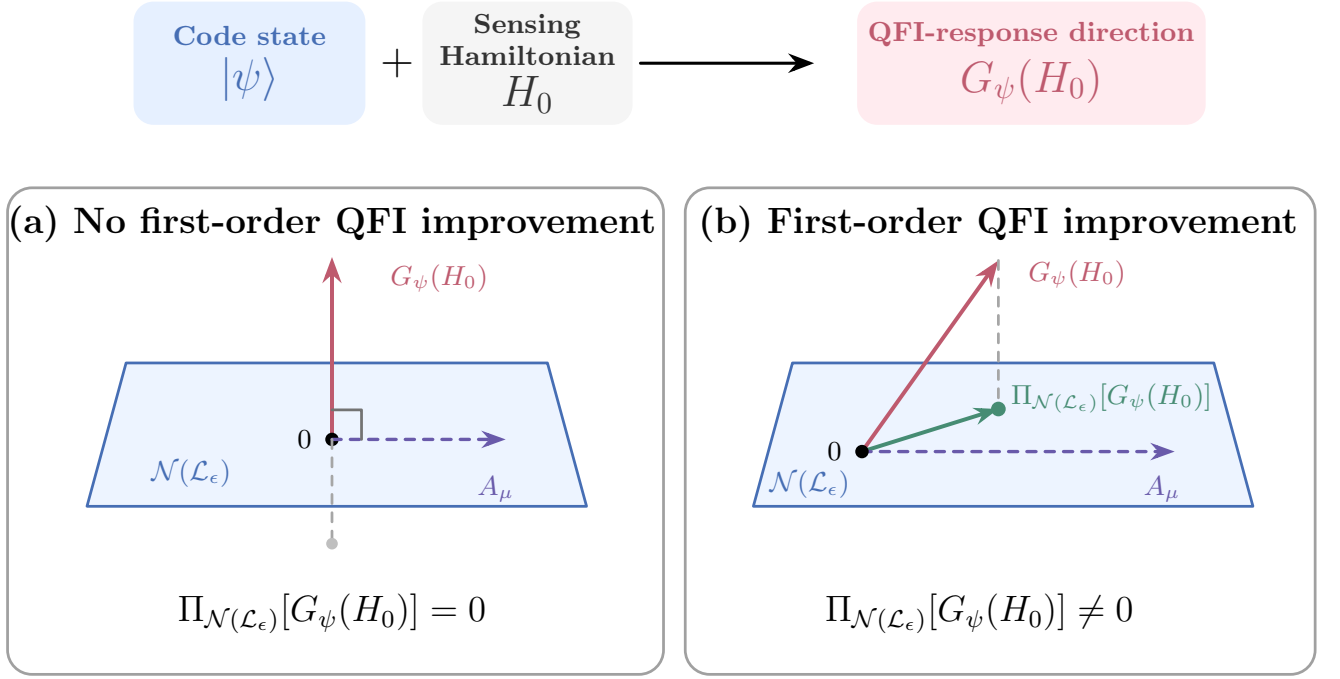
\begin{figure}[t]
\centering
\scalebox{1}{
\centering
\resizebox{0.98\textwidth}{!}{
\begin{tikzpicture}[
    >=Stealth,
    font=\small,
    line cap=round,
    line join=round,
    box/.style={
        draw,
        rounded corners=7pt,
        minimum height=1.45cm,
        align=center,
        inner sep=4pt,
        line width=0.9pt
    },
    bluebox/.style={
        box,
        draw=white,
        fill=macBlue!25,
        text=macBlueText
    },
    graybox/.style={
        box,
        draw=white,
        fill=macGray,
        text=black!80
    },
    redbox/.style={
        box,
        draw=white,
        fill=macPink!30,
        text=macPinkText
    },
    panel/.style={
        draw=macBorder,
        rounded corners=8pt,
        fill=white,
        line width=0.9pt
    },
    plane/.style={
        draw=macBlueText,
        fill=macBlue!18,
        line width=0.9pt
    },
    redvec/.style={
        -{Stealth[length=3mm,width=2.2mm]},
        line width=1.3pt,
        draw=macPinkText
    },
    greenvec/.style={
        -{Stealth[length=2.8mm,width=2mm]},
        line width=1.3pt,
        draw=macMintText
    },
    bluevec/.style={
        -{Stealth[length=2.8mm,width=2mm]},
        line width=1.1pt,
        dashed,
        draw=macLavText
    },
    projline/.style={
        dashed,
        draw=black!35,
        line width=0.9pt
    }
]

% ===================== Top row =====================
\coordinate (O4) at (3,9.2);
\node[bluebox, minimum width=2.7cm] (rho) at ($(O4)$)
{\textbf{Code state}\\[3pt]\Large $|\psi\rangle$};

\node at ($(O4)+(1.7,0)$) {\Large $+$};

\node[graybox, minimum width=2.45cm] (H) at ($(O4)+(3.2,0)$)
{\textbf{Sensing}\\\textbf{Hamiltonian}\\[3pt]\Large $H_0$};

%\node at ($(O4)+(5,0)$) {\Large $+$};

%\node[graybox, minimum width=3.55cm] (dH) at ($(O4)+(6,0)$)
%{\textbf{Centered Hamiltonian}\\[3pt]
%\Large $\Delta H_0$\\[-1pt]
%\scriptsize $\Delta H_0 = H_0 - \mathrm{Tr}(\rho H_0)\mathbb{I}$};

\draw[-{Stealth[length=3mm]}, line width=1pt] ($(O4)+(4.5,0)$) -- ($(O4)+(6.5,0)$);

\node[redbox, minimum width=4.0cm] (Gbox) at ($(O4)+(9,0)$)
{\textbf{QFI-response direction}\\[3pt]
\Large $G_\psi(H_0)$
};

% ===================== Left panel =====================
\node[panel, minimum width=7.55cm, minimum height=5.6cm, anchor=south west]
(Lpanel) at (0.2,2.15) {};

\begin{scope}[shift={(0.33,2.15)}]
    \node[font=\bfseries\large] at (3.62,5.2)
    {(a) No first-order QFI improvement};

%    \node[anchor=west] at (0.35,4.6) {Operator space};

    \coordinate (La) at (0.8,1.95);
    \coordinate (Lb) at (6.55,1.95);
    \coordinate (Lc) at (6.1,3.6);
    \coordinate (Ld) at (1.25,3.6);
    \draw[plane] (La)--(Lb)--(Lc)--(Ld)--cycle;

    \node[text=macBlueText, align=left] at (2,2.2)
    {$\mathcal{N}(\mathcal{L}_\epsilon)$\\[-1pt]
%     \scriptsize admissible symmetry\\[-1pt]
%     \scriptsize subspace
};

    \coordinate (O1) at (3.62,2.72);
    \coordinate (T1) at (3.62,4.82);

    \draw[redvec] (O1) -- (T1);
    \node[text=macPinkText, anchor=west] at (4.18,4.58)
    {$G_\psi(H_0)$};

    \fill[black] (O1) circle (2.2pt);
    \node[anchor=east, font=\bfseries] at ($(O1)+(-0.12,0)$) {$0$};

 	\draw[bluevec] (O1) -- ($(O1)+(2,0)$);
    \node[text=macLavText, anchor=west] at (5.18,2.15)
    {$A_\mu$};

    \draw[projline] (O1) -- ($(O1)+(0,-1.2)$);
    \fill[black!25] ($(O1)+(0,-1.2)$) circle (1.9pt);

    \draw[black!55, line width=0.9pt]
      ($(O1)+(0.34,0)$) --
      ($(O1)+(0.34,0.34)$) --
      ($(O1)+(0,0.34)$);

    \node[font=\large] at (3.62,0.72)
    {$\Pi_{\mathcal{N}(\mathcal{L}_\epsilon)}[G_\psi(H_0)] = 0$};
\end{scope}

% ===================== Right panel =====================
\node[panel, minimum width=7.25cm, minimum height=5.6cm, anchor=south west]
(Rpanel) at (8,2.15) {};

\begin{scope}[shift={(8,2.15)}]
    \node[font=\bfseries\large] at (3.62,5.2)
    {(b) First-order QFI improvement };

%    \node[anchor=west] at (0.35,4.6) {Operator space};

    \coordinate (Ra) at (0.8,1.95);
    \coordinate (Rb) at (6.55,1.95);
    \coordinate (Rc) at (6.1,3.6);
    \coordinate (Rd) at (1.25,3.6);
    \draw[plane] (Ra)--(Rb)--(Rc)--(Rd)--cycle;

    \node[text=macBlueText, align=left] at (1.45,2.2)
    {$\mathcal{N}(\mathcal{L}_\epsilon)$};

    \coordinate (O2) at (2.05,2.58);
    \coordinate (P2) at (3.62,3.07);
    \coordinate (T2) at (3.62,4.78);

    \draw[redvec] (O2) -- (T2);
    \node[text=macPinkText, anchor=west] at (3.85,4.62)
    {$G_\psi(H_0)$};

    \draw[projline] (T2) -- (P2);

    \draw[greenvec] (O2) -- (P2);
    \fill[macMintText] (P2) circle (2.4pt);
    \node[text=macMintText, anchor=west] at (3.65,3.2)
    {$\Pi_{\mathcal{N}(\mathcal{L}_\epsilon)}[G_\psi(H_0)]$};

    \draw[bluevec] (O2) -- (5.65,2.58);
    \node[text=macLavText, anchor=west] at (5.18,2.15)
    {$A_\mu$};

    \fill[black] (O2) circle (2.2pt);
    \node[anchor=east, font=\bfseries] at ($(O2)+(-0.12,0)$) {$0$};

    \node[font=\large] at (3.4,0.72)
    {$\Pi_{\mathcal{N}(\mathcal{L}_\epsilon)}[G_\psi(H_0)] \neq 0$};
\end{scope}

% ===================== Bottom criterion box =====================
%\node[
    draw=white,
    fill=SoftGray,
    rounded corners=7pt,
    line width=0.95pt,
    minimum width=9.0cm,
    minimum height=0.95cm,
    align=center,
    text=black,
    font=\large\bfseries
%] at (7.8,1)
%{Criterion:\quad
%$\Pi_{\mathcal{N}(\mathcal{L}_\epsilon)}[G_\psi(H_0)] \neq 0$};

\end{tikzpicture}}}

\caption{Geometric illustration of the first-order symmetry-optimization criterion.
The code state $|\psi\rangle$ and the sensing Hamiltonian $H_{0}$
determine the QFI-response direction $G_{\psi}(H_{0})$ in the operator
space. If $G_{\psi}(H_{0})$ is orthogonal to the subspace $\mathcal{N}(\mathcal{L}_{\epsilon})$
of admissible symmetry generators, its projection vanishes and no
first-order QFI improvement occurs. If $\Pi_{\mathcal{N}(\mathcal{L}_{\epsilon})}[G_{\psi}(H_{0})]\protect\neq0$,
there exists an admissible symmetry generator $A_{\mu}\in\mathcal{N}(\mathcal{L}_{\epsilon})$
that yields a nonzero linear response, so an appropriate symmetry
rotation generated by $A_{\mu}$ can increase the QFI.\label{SM:fig:First-order_criterion}}
\end{figure}

If the QFI $\hf_{\theta}$ can be increased by the symmetry optimization,
it requires that there exists at least one generator $A_{\mu}$ with
nonzero $q_{\mu}$ such that ${\rm Tr}\left[A_{\mu}G_{\psi}(H_{0})\right]\neq0$.
${\rm Tr}\left[A_{\mu}G_{\psi}(H_{0})\right]$ can be understood as
the coordinate of $G_{\psi}(H_{0})$ when it is projected onto $A_{\mu}$,
so this condition implies that $G_{\psi}(H_{0})$ is not orthogonal
to $\mathcal{N}(\mathcal{L}_{\epsilon})$ which is spanned by all
the symmetry generators. Therefore, the first-order HSO criterion
is
\begin{equation}
\Pi_{\mathcal{N}(\mathcal{L}_{\epsilon})}\left[G_{\psi}(H_{0})\right]\neq0,\label{SM:eq:HSO}
\end{equation}
where $\Pi_{\mathcal{N}(\mathcal{L}_{\epsilon})}$ denotes the projection
onto the admissible symmetry generator space $\mathcal{N}(\mathcal{L}_{\epsilon})$.
If this condition holds, there exists a symmetry generator $A_{\mu}\in\mathcal{N}(\mathcal{L}_{\epsilon})$
such that the QFI changes approximately linearly in $q_{\mu}$. Therefore,
choosing the sign of $q_{\mu}$ appropriately can increase the QFI.
The mechanism of enhancing QFI by the first-order HSO criterion is
illustrated in Fig. \ref{SM:fig:First-order_criterion}.

\subsection{Criterion for independent Pauli-error model}

For the independent Pauli-error model, the admissible symmetry generator
space is spanned by all single-qubit Pauli operators,
\begin{equation}
\mathcal{N}(\mathcal{L}_{\epsilon})={\rm span}_{\mathbb{R}}\{X_{j},Y_{j},Z_{j}:j=1,\ldots,n\},
\end{equation}
where the subscript $j$ denotes that the Pauli operator acts on the
$j$-th qubit and the span is over real scalars due to the Hermiticity
of symmetry generators. Choosing the normalized basis $A_{j,\alpha}=\sigma^{\alpha}_{j}$,
$\alpha=x,y,z$, with the convention $\frac{1}{2^{n}}{\rm Tr}\left(\sigma^{\alpha}_{j}\sigma^{\beta}_{k}\right)=\delta_{jk}\delta_{\alpha\beta}$,
the projection operator becomes
\begin{equation}
\Pi_{{\rm 1body}}(X)=\frac{1}{2^{n}}\sum^{n}_{j=1}\sum_{\alpha=x,y,z}{\rm Tr}\left(\sigma^{\alpha}_{j}X\right)\sigma^{\alpha}_{j}.
\end{equation}
Consequently, in this case the first-order HSO condition is
\begin{equation}
\Pi_{{\rm 1body}}\left[H_{0},\rho_{\psi}\Delta H_{0}+\Delta H_{0}\rho_{\psi}\right]\neq0,
\end{equation}
or, equivalently, $\left[H_{0},\rho_{\psi}\Delta H_{0}+\Delta H_{0}\rho_{\psi}\right]$
has at least one nonzero one-body Pauli component.

\subsection{Second-order HSO criterion\label{SM:subsec:Second-order-HSO-criterion}}

When the first-order variation vanishes, the QFI is governed by the
Hessian matrix with respect to the symmetry generators.

The Hessian matrix is given by
\begin{equation}
\mathcal{H}_{\mu\nu}=\left.\frac{\partial^{2}\hf_{\theta}(\boldsymbol{q},\psi)}{\partial q_{\mu}\partial q_{\nu}}\right|_{\boldsymbol{q}=\boldsymbol{0}}.
\end{equation}
To compute it, we define
\begin{equation}
D_{\mu}=\left.\frac{\partial K(\boldsymbol{q})}{\partial q_{\mu}}\right|_{\boldsymbol{q}=\boldsymbol{0}}=\i[A_{\mu},H_{0}],
\end{equation}
and
\begin{equation}
E_{\mu\nu}=\left.\frac{\partial^{2}K(\boldsymbol{q})}{\partial q_{\mu}\partial q_{\nu}}\right|_{\boldsymbol{q}=\boldsymbol{0}}=-\frac{1}{2}\left([A_{\mu},[A_{\nu},H_{0}]]+[A_{\nu},[A_{\mu},H_{0}]]\right).
\end{equation}
The symmetrized form of $E_{\mu\nu}$ appears because the mixed derivatives
with respect to $q_{\mu}$ and $q_{\nu}$ commute. Since $\hf_{\theta}(\boldsymbol{q},\psi)=4t^{2}{\rm Var}_{\psi}[K(\boldsymbol{q})]$,
a direct differentiation gives
\begin{equation}
\mathcal{H}_{\mu\nu}=8t^{2}\left[{\rm Cov}_{\psi}(D_{\mu},D_{\nu})+{\rm Cov}_{\psi}(H_{0},E_{\mu\nu})\right].
\end{equation}

When the first-order variation vanishes, if the QFI can be increased
by the symmetry optimization, there must exist at least one symmetry
generator $A_{\boldsymbol{c}}=\sum_{\mu}c_{\mu}A_{\mu}$ such that
$\boldsymbol{c}^{T}\mathcal{H}\boldsymbol{c}>0$, where $\boldsymbol{c}$
is the vector of the coefficients $c_{\mu}$. Therefore, a second-order
condition for symmetry optimization is
\begin{equation}
\lambda_{\max}(\mathcal{H})>0,
\end{equation}
i.e., the Hessian matrix $\mathcal{H}$ must not be negative semi-definite.

\subsection{Criterion for dark state}

When the initial code state is dark with respect to $H_{0}$, i.e.,
$H_{0}|\psi\rangle=\lambda|\psi\rangle$, a particularly simple expression
can be obtained for the second-order criterion.

In this case, $F_{\theta}(\boldsymbol{0},\psi)=0$ and the first-order
derivative vanishes automatically. Moreover, ${\rm Cov}_{\psi}(H_{0},E_{\mu\nu})=0$,
so the Hessian matrix is reduced to the Gram matrix,
\begin{equation}
\mathcal{H}_{\mu\nu}=8t^{2}{\rm Cov}_{\psi}(D_{\mu},D_{\nu})=8t^{2}{\rm Re}\langle D_{\mu}\psi|\Pi^{\perp}_{\psi}|D_{\nu}\psi\rangle,
\end{equation}
where $\Pi^{\perp}_{\psi}=I-|\psi\rangle\langle\psi|$ is the projector
onto the subspace orthogonal to the initial code state $|\psi\rangle$.

Equivalently, for any symmetry generator $A_{\boldsymbol{c}}=\sum_{\mu}c_{\mu}A_{\mu}\in\mathcal{N}(\mathcal{L}_{\epsilon})$,
\begin{equation}
\boldsymbol{c}^{T}\mathcal{H}\boldsymbol{c}=8t^{2}\left|\Pi^{\perp}_{\psi}[A_{\boldsymbol{c}},H_{0}]|\psi\rangle\right|^{2}\ge0.
\end{equation}
Thus, for a dark code state, the second-order HSO condition is reduced
to

\begin{equation}
\exists A\in\mathcal{N}(\mathcal{L}_{\epsilon}),\quad\Pi^{\perp}_{\psi}[A,H_{0}]|\psi\rangle\neq0,
\end{equation}
implying that the initial code state $|\psi\rangle$ must not be an
eigenstate of $[A,H_{0}]$ for at least one admissible symmetry generator
$A$.

\subsection{Criterion for stabilizer-sum Hamiltonian}

A particularly interesting example of dark state is that the initial
state is a stabilizer code state and the Hamiltonian is a sum of the
stabilizers. In this case, the symmetry optimization condition can
be significantly simplified.

Let $\mathcal{C}$ be the code space of a stabilizer code with stabilizer
generators $\{g_{j}\}^{m}_{j=1}$, where $m=n-l$. Consider a stabilizer-sum
Hamiltonian, $H(\theta)=\theta H_{0}$, where
\begin{equation}
H_{0}=-\sum^{m}_{j=1}g_{j}.
\end{equation}
For any code state $|\psi\rangle\in\mathcal{C}$, one has $g_{j}|\psi\rangle=|\psi\rangle$,
$j=1,\ldots,m$, and therefore $H_{0}|\psi\rangle=-m|\psi\rangle$.
Thus every code state is a dark state for the unrotated stabilizer-sum
Hamiltonian, and the unrotated QFI vanishes,
\begin{equation}
\hf_{\theta}(\boldsymbol{0},\psi)=0.
\end{equation}

The general second-order dark-state HSO criterion derived above states
that QFI can be increased by an infinitesimal admissible symmetry
transformation if and only if there exists $A\in\mathcal{N}(\mathcal{L}_{\epsilon})$
such that $\Pi^{\perp}_{\psi}[A,H_{0}]|\psi\rangle\neq0$. For a stabilizer-sum
Hamiltonian, this condition can be evaluated explicitly. We have
\begin{equation}
[A,H_{0}]|\psi\rangle=-\sum^{m}_{j=1}[A,g_{j}]|\psi\rangle.
\end{equation}
For each stabilizer generator $g_{j}$, if $[A,g_{j}]=0$, the corresponding
term vanishes.

For a Pauli symmetry generator $A$, if $\{A,g_{j}\}=0$, then $[A,g_{j}]=2Ag_{j}$,
and hence $[A,g_{j}]|\psi\rangle=2A|\psi\rangle$. Let $\mathcal{J}_{A}=\{g_{j}\,|\,\{A,g_{j}\}=0\}$
be the set of stabilizer generators that anticommute with $A$, and
let $r_{A}=|\mathcal{J}_{A}|$ be the number of such stabilizer generators.
Then,
\begin{equation}
[A,H_{0}]|\psi\rangle=-2r_{A}A|\psi\rangle.
\end{equation}
Consequently,
\begin{equation}
\Pi^{\perp}_{\psi}[A,H_{0}]|\psi\rangle=-2r_{A}\Pi^{\perp}_{\psi}A|\psi\rangle,
\end{equation}
where $A|\psi\rangle$ lies in the stabilizer-syndrome subspace obtained
by flipping precisely the stabilizers in $\mathcal{J}_{A}$. If $r_{A}>0$,
this syndrome subspace is different from the original code space and
is therefore orthogonal to $|\psi\rangle$. Hence,
\begin{equation}
\Pi^{\perp}_{\psi}A|\psi\rangle=A|\psi\rangle\neq0.
\end{equation}
It follows that the dark-state HSO condition is reduced to
\begin{equation}
\exists A\in\mathcal{N}(\mathcal{L}_{\epsilon}),\quad r_{A}>0.
\end{equation}
Equivalently, there must exist an admissible symmetry generator $A$
that anticommutes with at least one stabilizer generator $g_{j}$,
i.e., $\{A,g_{j}\}=0$. Under this condition, the Hessian of the QFI
has a positive direction. Indeed, with $U_{A}(q)=e^{-\i qA}$, the
QFI can be derived as
\begin{equation}
\begin{aligned}\hf_{\theta}(q,\psi)= & 4t^{2}{\rm Var}_{\psi}\left(e^{\i qA}H_{0}e^{-\i qA}\right)\\
= & 4t^{2}q^{2}\left|\Pi^{\perp}_{\psi}[A,H_{0}]|\psi\rangle\right|^{2}+O(q^{3})\\
= & 16t^{2}q^{2}r^{2}_{A}{\rm Var}_{\psi}(A)+O(q^{3}),
\end{aligned}
\end{equation}
and therefore $\hf_{\theta}>0$ for sufficiently small nonzero $q$
whenever $r_{A}>0$.

When the errors are independent, the admissible symmetry generators
are all one-body Pauli operators, so the above condition can be further
simplified to
\begin{equation}
\exists i,\alpha,j,\quad\{\sigma^{\alpha}_{i},g_{j}\}=0.
\end{equation}
For a stabilizer-sum Hamiltonian, this condition can generally be
satisfied, as the stabilizer generators are all Pauli strings and
one can always find a single-qubit Pauli operator to anticommute with
at least one stabilizer in the Hamiltonian. Therefore, a stabilizer
code can generally be optimized for the associated stabilizer-sum
Hamiltonian to increase the QFI under independent errors.

Finally, it is worth noting that we have used an equivalent fixed-code
Hamiltonian-rotation representation instead of the code-rotation picture
in the above derivation. For a code state $|\psi\rangle$ and a noise-symmetry
transformation $U$, the QFI of the rotated code under the original
Hamiltonian $H_{0}$ is determined by
\begin{equation}
\operatorname{Var}_{U|\psi\rangle}(H_{0})=\operatorname{Var}_{|\psi\rangle}(U^{\dagger}H_{0}U).
\end{equation}
Therefore, rotating the code while keeping the signal Hamiltonian
fixed is mathematically equivalent to keeping the code fixed and rotating
the Hamiltonian as
\begin{equation}
H_{U}=U^{\dagger}H_{0}U.
\end{equation}
The two pictures are mathematically equivalent, although they correspond
to different physical implementations: code rotation requires implementing
a transformation in the encoding circuit, whereas Hamiltonian rotation
requires engineering the sensing dynamics.

\section{Bound of quantum Fisher information}

Now, we consider the symmetry optimization of the stabilizer code
$\llbracket n,l,d\rrbracket$ for a stabilizer-sum Hamiltonian, which
meets the second-order HSO condition derived in Sec. \ref{SM:subsec:Second-order-HSO-criterion}
and allows effective symmetry optimization,
\begin{equation}
H(\theta)=\theta H_{0},\;H_{0}=-\sum^{n-l}_{i=1}g_{i},\label{SM:eq:smHami}
\end{equation}
where $\{g_{i}\}$ are the generators of the underlying stabilizer
group and $\theta$ is the unknown parameter to estimate.

Suppose one can identify multiple commuting Pauli operators $\{A_{j}\}$
such that each $A_{j}$ anticommutes with a subset of the stabilizer
generators $\{g_{i}\}$ and commutes with the rest, and these subsets
are disjoint. One can apply each $\exp(-\text{i}q_{j}A_{j})$ to the
initial code state $|\psi_{0}\rangle$ to enhance the QFI while keeping
the correctable set of noise invariant, and the symmetry transformation
is
\begin{equation}
\widetilde{U}=\prod^{k}_{j=1}\text{exp}(-\text{i}q_{j}A_{j}),
\end{equation}
where $k$ is the number of generators $\{A_{j}\}$. Assume each of
the generators $\{A_{j}\}$ anticommutes with $r_{j}$ different stabilizer
generators, respectively, as shown in Fig. \ref{SM:fig:generatorof control}.
The total number of the stabilizer generators that anticommute with
one of $\{A_{j}\}$ does not exceed the number of $\{g_{i}\}$, i.e.,
$\sum^{k}_{j=1}r_{j}\leq n-l$.

Let $\left|-m\right\rangle $ denote one of the ground states of the
stabilizer-sum Hamiltonian corresponding to its smallest eigenvalue
$-m$. Since the code space is the ground space of the Hamiltonian,
the initial state of the system can be taken to be $\left|-m\right\rangle $.
Consequently, the QFI for this system after the symmetry optimization
by $\widetilde{U}$ can be expressed as
\begin{align}
\mathcal{F}_{\theta} & =4t^{2}\left(\left\langle -m\left|\widetilde{U}^{\dagger}\left(\frac{\partial H(\theta)}{\partial\theta}\right)^{2}\widetilde{U}\right|-m\right\rangle -\left\langle -m\left|\widetilde{U}^{\dagger}\left(\frac{\partial H(\theta)}{\partial\theta}\right)\widetilde{U}\right|-m\right\rangle ^{2}\right).\label{SM:eq:sf}
\end{align}
For simplicity, we define
\begin{equation}
\epsilon_{ij}=\begin{cases}
1 & \text{if }[A_{j},g_{i}]=0,\\
-1 & \text{if }\{A_{j},g_{i}\}=0,
\end{cases}
\end{equation}
for any $i\in\{1,2,\cdots,n-l\}$ and $j\in\{1,2,\cdots,k\}$. We
have
\begin{align}
\left\langle \widetilde{U}^{\dagger}\left(\frac{\partial H(\theta)}{\partial\theta}\right)\widetilde{U}\right\rangle  & =\left\langle -m\left|\left(\prod^{k}_{j=1}\text{exp}(-\text{i}q_{j}A_{j})\right)\sum^{n-l}_{i=1}(-g_{i})\left(\prod^{k}_{j=1}\text{exp}(\text{i}q_{j}A_{j})\right)^{\dagger}\right|-m\right\rangle \nonumber \\
 & =-\left\langle -m\left|\left(\prod^{k}_{j=1}\text{exp}(-\text{i}q_{j}A_{j})\right)\sum^{n-l}_{i=1}\left(\prod^{k}_{j=1}\text{exp}(\text{i}\epsilon_{ij}q_{j}A_{j})\right)^{\dagger}g_{i}\right|-m\right\rangle \nonumber \\
 & =-\left\langle -m\left|\left(\prod^{k}_{j=1}\text{exp}(-\text{i}q_{j}A_{j})\right)\sum^{k}_{i=1}r_{i}\left(\prod^{k}_{j=1}\text{exp}(\text{i}(-1)^{\delta_{ij}}q_{j}A_{j})\right)^{\dagger}\right|-m\right\rangle \nonumber \\
 & =-\left\langle -m\left|\sum^{k}_{i=1}r_{i}\left(\prod^{k}_{j=1}\text{exp}(-\text{i}q_{j}A_{j})\right)\text{exp}(-2\text{i}q_{i}A_{i})\left(\prod^{k}_{j=1,\neq i}\text{exp}(\text{i}q_{j}A_{j})\right)^{\dagger}\right|-m\right\rangle \nonumber \\
 & =-\left\langle -m\left|\sum^{k}_{i=1}r_{i}\left[\text{cos}(2q_{i})\mathbb{I}-\text{i}\text{sin}(2q_{i})A_{i}\right]\right|-m\right\rangle ,\label{SM:eq:suhu}
\end{align}
where $\delta_{ij}$ is the Kronecker delta function, i.e., 
\begin{figure}[t]
\centering
\scalebox{1.2}{
\begin{tikzpicture}

% Define styles
\tikzset{
    box/.style={rectangle, draw, fill=red!30, rounded corners},
    brace/.style={decorate, decoration={brace, amplitude=10pt}},
    label/.style={font=\small}
}

% Draw boxes
\node[box] (A1) at (0, 0) {${g_1, g_2, \cdots, g_{r_1}}$};
\node[box] (A2) at (2.7, 0) {${g_{r_1+1}, \cdots, g_{r_1+r_2}}$};
\node[box] (Ak) at (6.8, -0.1) {${g_{\sum_{j=1}^{k-1} r_j + 1}, \cdots, g_{\sum_{j=1}^k r_j}}$};

% Draw braces
\draw[brace]    (A1.south east)-- (A1.south west) node[midway, left=5pt, label] {$ $};
\draw[brace]      (A2.south east)--  (A2.south west) node[midway, left=5pt, label] {$ $};
\draw[brace]      (Ak.south east)  --(Ak.south west) node[midway, left=5pt, label] {$ $};

% Labels under boxes
\node at (0, -1.2) {$A_1$};
\node at (2.7, -1.2) {$A_2$};
\node at (6.8, -1.33) {$A_k$};

\node at (0, -0.8) {$r_1$};
\node at (2.7, -0.8) {$r_2$};
\node at (6.8, -0.99) {$r_k$};

\node at (4.5, 0) {$\cdots$};
\node at (4.5, -0.9) {$\cdots$};

% Braces at the top
\draw [decorate,decoration={brace,mirror,amplitude=15pt},yshift=15pt]  (Ak.north east)--(A1.north west)  node [midway, yshift=20pt] {$\sum r_j$ stabilizer generators};

	\node at (-1.64, 0) {$\{g_i\}$};
%	\node at (0, 0) {000};

\end{tikzpicture}}

\caption{Each $A_{j}$ anticommutes with $r_{j}$ stabilizer generators only.
All $k$ symmetry generators $\{A_{j}\}$ anticommute with $\sum^{k}_{j=1}r_{j}$
stabilizer generators jointly.\label{SM:fig:generatorof control}}
\end{figure}
\begin{equation}
\delta_{ij}=\begin{cases}
1 & \text{if }i=j,\\
0 & \text{if }i\ne j,
\end{cases}
\end{equation}
and a constant term $c\mathbb{I}$ has been dropped from the third
line as it does not affect the variance of $\widetilde{U}^{\dagger}\left(\frac{\partial H(\theta)}{\partial\theta}\right)\widetilde{U}$
in the QFI \eqref{SM:eq:sf}. $c$ is the number of stabilizer generators
that do not anticommute with any $A_{j}$, i.e.,
\begin{equation}
c=n-l-\sum^{k}_{i=1}r_{i}.
\end{equation}
The third equal sign of Eq. \eqref{SM:eq:suhu} comes by observing that
each $g_{i}$ anticommutes with at most one $A_{j}$. So, the second
term of Eq. \eqref{SM:eq:sf} turns out to be
\begin{align}
\left\langle \widetilde{U}^{\dagger}\left(\frac{\partial H(\theta)}{\partial\theta}\right)\widetilde{U}\right\rangle ^{2} & =\left\langle -m\left|\sum^{k}_{i=1}r_{i}\left[\text{cos}(2q_{i})\mathbb{I}-\text{i}\text{sin}(2q_{i})A_{i}\right]\right|-m\right\rangle ^{2}=\left(\sum^{k}_{i=1}r_{i}\text{cos}(2q_{i})\right)^{2},\label{SM:eq:1-1}
\end{align}
where we have used $\left\langle -m\left|A_{i}\right|-m\right\rangle =0$
due to $\{A_{i},g_{j}\}=0$ for some stabilizer generator $g_{j}$.

Similar to Eq. \eqref{SM:eq:suhu}, the first term of Eq. \eqref{SM:eq:sf}
can be rewritten as
\begin{align}
\left\langle \widetilde{U}^{\dagger}\left(\frac{\partial H(\theta)}{\partial\theta}\right)^{2}\widetilde{U}\right\rangle  & =\left\langle \left(\widetilde{U}^{\dagger}\left(\frac{\partial H(\theta)}{\partial\theta}\right)\widetilde{U}\right)\left(\widetilde{U}^{\dagger}\left(\frac{\partial H(\theta)}{\partial\theta}\right)\widetilde{U}\right)\right\rangle \\
 & =\left\langle -m\left|\left\{ \sum^{k}_{i=1}r_{i}\left[\text{cos}(2q_{i})\mathbb{I}+\text{i}\text{sin}(2q_{i})A_{i}\right]\right\} \left\{ \sum^{k}_{i=1}r_{i}\left[\text{cos}(2q_{i})\mathbb{I}-\text{i}\text{sin}(2q_{i})A_{i}\right]\right\} \right|-m\right\rangle \\
 & =\left(\sum^{k}_{i=1}r_{i}\text{cos}(2q_{i})\right)^{2}+\left(\sum^{k}_{i,j}r_{i}r_{j}\text{sin}(2q_{i})\text{sin}(2q_{j})\left\langle -m\left|A_{i}A_{j}\right|-m\right\rangle \right)\label{SM:eq:suh2u1}\\
 & =\left(\sum^{k}_{i=1}r_{i}\text{cos}(2q_{i})\right)^{2}+\left(\sum^{k}_{i,j}r_{i}r_{j}\text{sin}(2q_{i})\text{sin}(2q_{j})\delta_{ij}\right)\label{SM:eq:suh2u2}\\
 & =\left(\sum^{k}_{i=1}r_{i}\text{cos}(2q_{i})\right)^{2}+\sum^{k}_{i}r^{2}_{i}\text{sin}^{2}(2q_{i}),\label{SM:eq:1-2}
\end{align}
where a constant term $c\mathbb{I}$ has also been dropped from both
braces in the second line as it does not affect the variance of $\widetilde{U}^{\dagger}\left(\frac{\partial H(\theta)}{\partial\theta}\right)\widetilde{U}$.

The step from Eq.~\eqref{SM:eq:suh2u1} to Eq.~\eqref{SM:eq:suh2u2} relies
on the identity
\begin{equation}
\left\langle -m\left|A_{i}A_{j}\right|-m\right\rangle =\delta_{ij},\label{SM:eq:mAAm}
\end{equation}
which we now verify. Let $\hs_{i}$ denote the subset of stabilizers
that anticommute with $A_{i}$. By construction, the subsets $\hs_{j}$
are pairwise distinct and disjoint. Then $A_{i}|-m\rangle$ lies in
a stabilizer syndrome subspace in which the eigenvalues of the stabilizers
in $\hs_{i}$ are flipped, while all other stabilizer eigenvalues
remain unchanged. For $i\neq j$, even if $r_{i}=r_{j}$ and the states
$A_{i}|-m\rangle$ and $A_{j}|-m\rangle$ may lie in the same eigenspace
of $H(\theta)$, they belong to distinct stabilizer syndrome subspaces,
since $A_{i}$ and $A_{j}$ flip different subsets $\mathcal{S}_{i}$
and $\mathcal{S}_{j}$ of the stabilizers. There exists at least one
stabilizer $g$ belonging to $\hs_{i}$ but not $\hs_{j}$, or vice
versa. The eigenvalue of $g$ is flipped by $A_{i}$ but not by $A_{j}$.
Hence $A_{i}|-m\rangle$ and $A_{j}|-m\rangle$ belong to different
eigenspaces of the Hermitian stabilizer of $g$, and are therefore
orthogonal:
\[
\left\langle -m\left|A_{i}A_{j}\right|-m\right\rangle =0,\qquad i\ne j.
\]
For $i=j$ and $A^{2}_{i}=I$, we have
\[
\left\langle -m\left|A_{i}A_{j}\right|-m\right\rangle =1,\qquad i=j.
\]
Thus, Eq.~\eqref{SM:eq:mAAm} is obtained.

Substituting Eq.~\eqref{SM:eq:1-1} and \eqref{SM:eq:1-2} into Eq.~\eqref{SM:eq:sf},
we can obtain a simplified expression of QFI in Eq.~\eqref{SM:eq:sf},\textcolor{black}{
\begin{equation}
\mathcal{F}_{\theta}=4t^{2}\sum^{k}_{i=1}r^{2}_{i}\text{sin}^{2}(2q_{i}).
\end{equation}
}Therefore, setting $q_{i}=\pi/4$ for all $i$ and using the Cauchy--Schwarz
inequality, one can obtain a lower bound for the maximum of QFI,
\begin{equation}
(\mathcal{F}_{\theta})_{\max}=4t^{2}\sum^{k}_{i=1}r^{2}_{i}\ge\frac{4t^{2}}{k}\left(\sum^{k}_{i=1}r_{i}\right)^{2}.\label{SM:eq:sbound}
\end{equation}

It indicates that the more the stabilizer generators anticommute with
one of the symmetry generators $\{A_{j}\}$, the larger the QFI becomes.
In the scenario that each $A_{j}$ anticommutes with a distinct subset
of the stabilizer generators $\{g_{i}\}$ and these subsets are disjoint
while together covering all $\{g_{i}\}$, i.e.,
\begin{equation}
\sum^{k}_{i=1}r_{i}=n-l,
\end{equation}
the maximum of QFI reaches the largest lower bound,
\begin{equation}
(\mathcal{F}_{\theta})_{\max}\ge4t^{2}\frac{(n-l)^{2}}{k}.\label{SM:eq:fmax}
\end{equation}
For some stabilizer codes, such as the surface code discussed in the
next section, the number of $\{A_{j}\}$ is proportional to the total
number of physical qubits $n$, i.e., $k=O(n).$ The maximum of QFI
of these systems exhibits at least SQL scaling, i.e., $(\mathcal{F}_{\theta})_{\max}=O(n).$

\section{Achieving SQL with surface code}

In this section, we take the surface code \cite{SM-PhysRevA.86.032324,SM-Horsman2011SurfaceCQ,SM-Cleland2022AnIT}
as an example to show that the QFI can be increased to the SQL scaling
with respect to the number of data qubits through symmetry optimization
of the code space. The surface code is a paradigmatic stabilizer QEC
code. The idea behind surface code is to encode information in the
homological degrees of freedom of a two-dimensional qubit lattice,
which makes it a powerful architecture for fault-tolerant quantum
computation due to its robustness against local noise. The effectiveness
of surface codes has been demonstrated in various experiments \cite{SM-Krinner2022,SM-Acharya2025,SM-zhaoRealizationErrorCorrectingSurface2022a,SM-PhysRevLett.107.240501,SM-Mariantoni2011ImplementingTQ,SM-Krinner2022,SM-Bluvstein2022}.
\begin{figure}
\subfloat[Stabilizer generators of surface code\label{SM:fig:Surface-code-1}]{\includegraphics[totalheight=6.5cm]{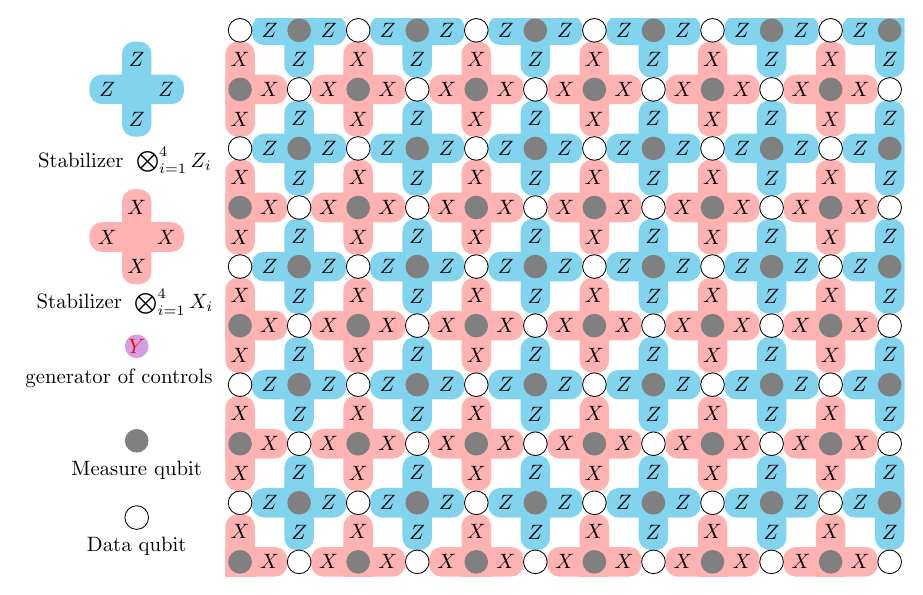}

}\subfloat[Controls on surface code to anticommute with stabilizer generators]{\includegraphics[totalheight=6.5cm]{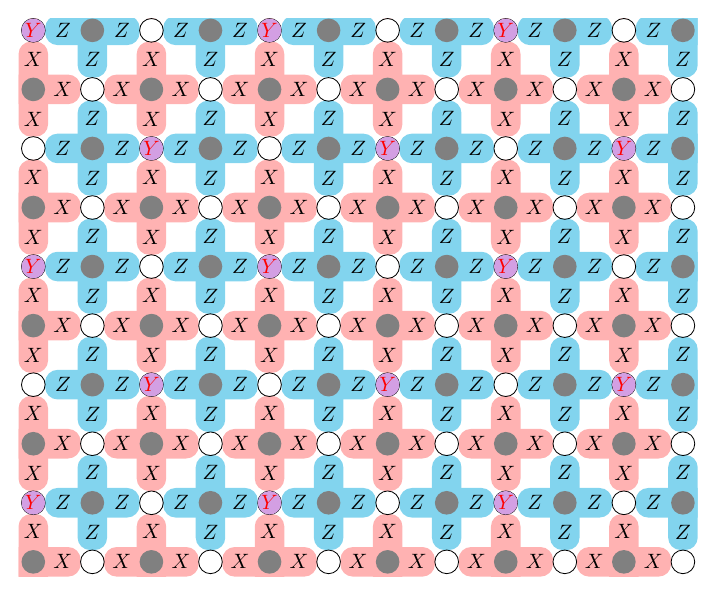}

\label{SM:Sfig_surface_code_withcontrol}}

\caption{Illustration of (a) the stabilizer generators of the surface code
and (b) the local controls used to anticommute with all stabilizer
generators. The physical qubits are either data qubits (open circles)
or measurement qubits (black solid circles). One only needs to apply
single-qubit controls on the qubits marked with purple solid circles,
so that SQL can be achieved by surface codes with stabilizer-sum Hamiltonian.}
\end{figure}

Suppose that the surface code is defined on a two-dimensional $N\times N$
lattice of $N^{2}$ physical qubits. These qubits are either data
qubits, in which the computational quantum states are stored, or measurement
qubits, which are used to stabilize and manipulate the quantum state
of the data qubits. The parameter-dependent Hamiltonian acts on the
data qubits only, and the measurement qubits in the figure are only
to indicate the stabilizer structure. Thus, $n=N^{2}/2.$ The stabilizer
group of surface codes is generated by two types of local operators,
\begin{equation}
S^{(x)}_{f}\coloneqq\bigotimes_{i\in G_{f}}X_{i},\quad S^{(z)}_{f'}\coloneqq\bigotimes_{i\in G_{f'}}Z_{i},
\end{equation}
where $G_{f}$ and $G_{f'}$ are the sets including four vertices
closest to the $f$-th and $f'$-th measurement qubits,  respectively.
And $S^{(x)}_{f}$and $S^{(z)}_{f'}$ only act non-trivially on the
four data qubits surrounding the $f$-th and $f'$-th measurement
qubit respectively, as shown in Fig. \ref{SM:fig:Surface-code-1}.

We now consider a stabilizer-sum Hamiltonian, 
\begin{equation}
H(\theta)=-\theta\left(\sum_{f}S^{(x)}_{f}+\sum_{f'}S^{(z)}_{f'}\right).
\end{equation}
As the logical space of the surface code is the ground space of $H(\theta)$,
an initial codeword has zero QFI with respect to $\theta$ under this
Hamiltonian. To achieve a nonzero QFI, the code space must be optimized
by exploiting the symmetry of the noise, so as to drive the qubits
outside the ground space of the stabilizer-sum Hamiltonian. As plotted
in Fig. \ref{SM:Sfig_surface_code_withcontrol}, a key feature of the
surface code is that each stabilizer involves only $O(1)$ qubits,
and each qubit participates in only a constant number of stabilizers.
Consequently, a single local generator $A_{j}$ can anticommute with
at most $O(1)$ stabilizers. So, the total number $k$ of $A_{j}$'s
that anticommute with all generators scales as $O(n)$, i.e., $k=O(n)$.
Thus, the bound of QFI in Eq. \eqref{SM:eq:sbound} scales as $O(n)$,
\begin{equation}
(\mathcal{F}_{\theta})_{\max}\ge\frac{4t^{2}(n-l)^{2}}{k}=O(n),
\end{equation}
which is the SQL scaling with respect to the number of physical qubits.

\section{Another example to surpass SQL---TDC code}

For some systems discussed in the main text, the bound of QFI can
exceed the SQL of physical qubits and even reach the Heisenberg limit.
Here we present another example in which the bound of QFI can exceed
the SQL. This construction demonstrates that the symmetry optimization
can produce intermediate precision scalings when the symmetry generators
act on a growing number of stabilizers but not on all of them.

Consider an operator quantum error-correcting code defined on a two-dimensional
$N\times N$ lattice of $n$ physical qubits, where $N$ is even and
$n=N^{2}$. This code employs a subsystem structure in its error recovery
procedure and shares lattice-based features with the surface codes
\cite{SM-PhysRevA.86.032324,SM-Horsman2011SurfaceCQ,SM-Cleland2022AnIT}. For
convenience, we refer to it as two-dimensional cruciform code (TDC
code).

\begin{figure}
\includegraphics[width=10cm]{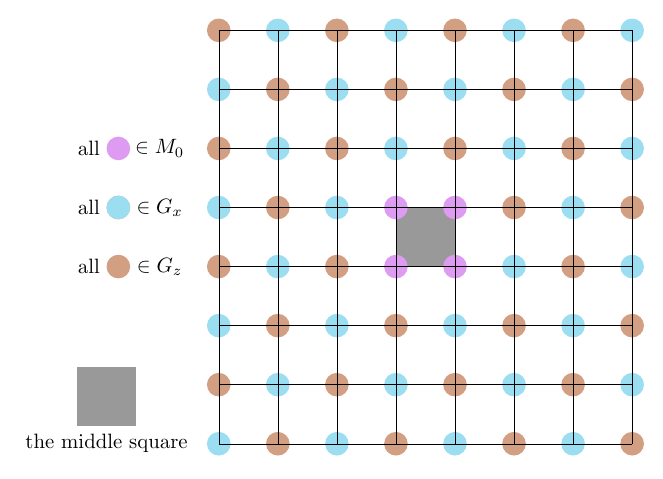}

\caption{\label{SM:fig:TDC-1}The lattice is partitioned into three disjoint sets:
$G_{x}$ (blue), $G_{z}$ (red), and $M_{0}$\LyXZeroWidthSpace{} (brown).
The gray square denotes the central region.}
\end{figure}

\begin{figure*}
\noindent\begin{minipage}[t]{1\columnwidth}%
\subfloat[Stabilizer $S^{X}_{(3,5)}$$\centering$]{\includegraphics[width=4.5cm]{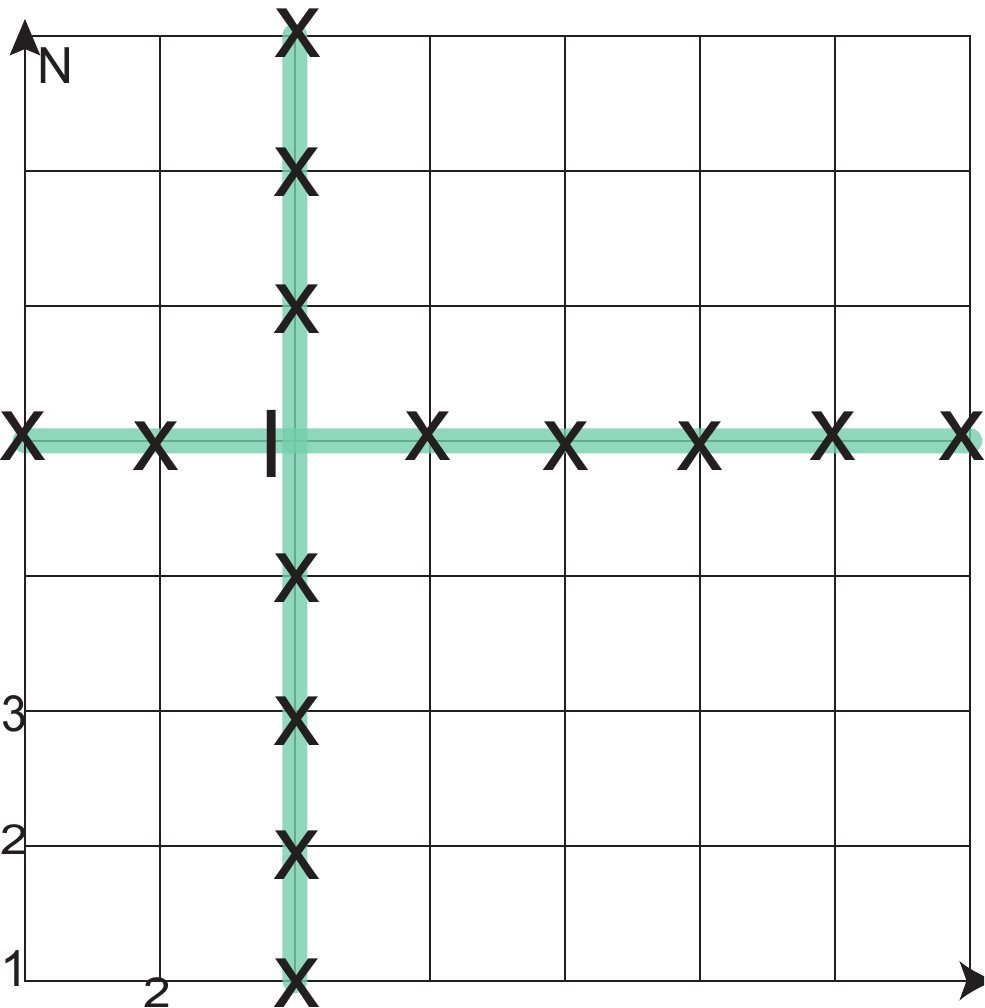}

}\subfloat[Stabilizer $S^{Z}_{(5,3)}$$\centering$]{\includegraphics[width=4.5cm]{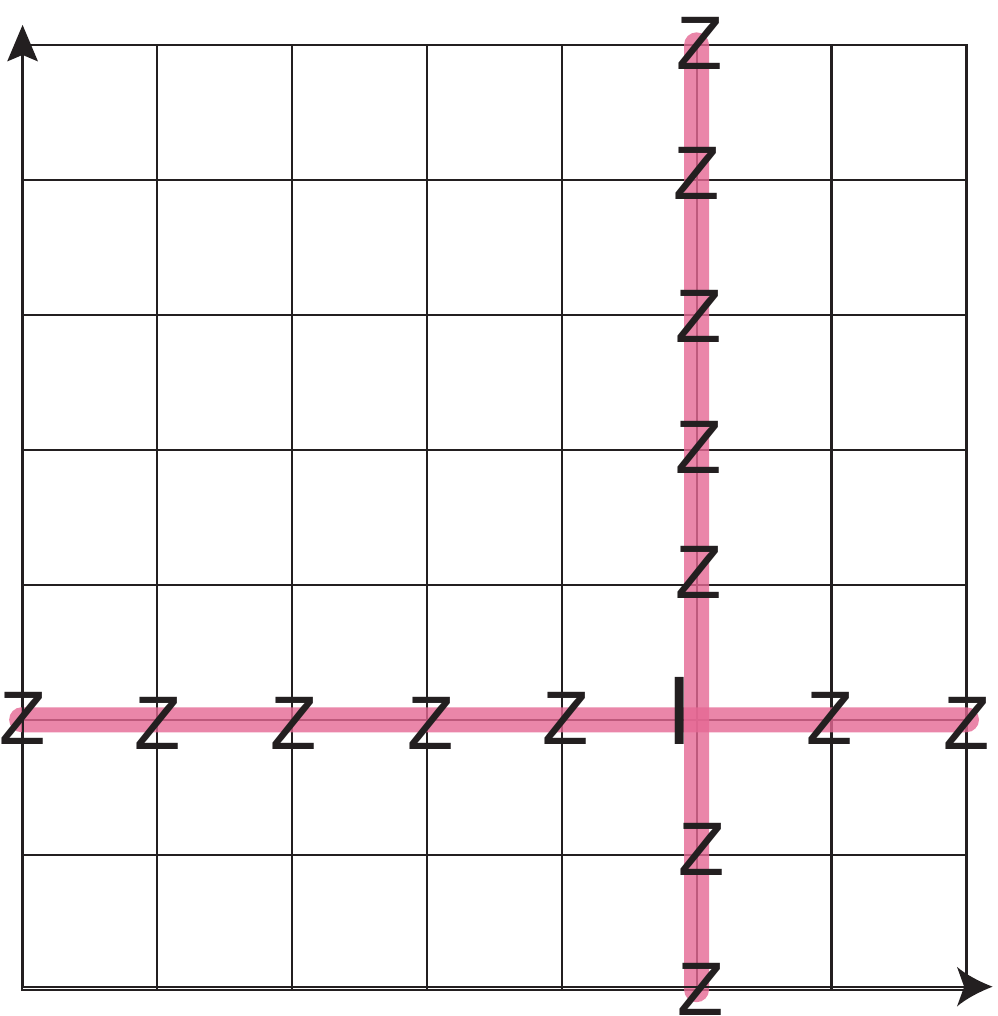}

}\subfloat[Logical operators $\overline{X}_{(N/2,N/2)}$ and $\overline{Z}_{(N/2+1,N/2+1)}$
$\centering$]{\includegraphics[width=4.5cm]{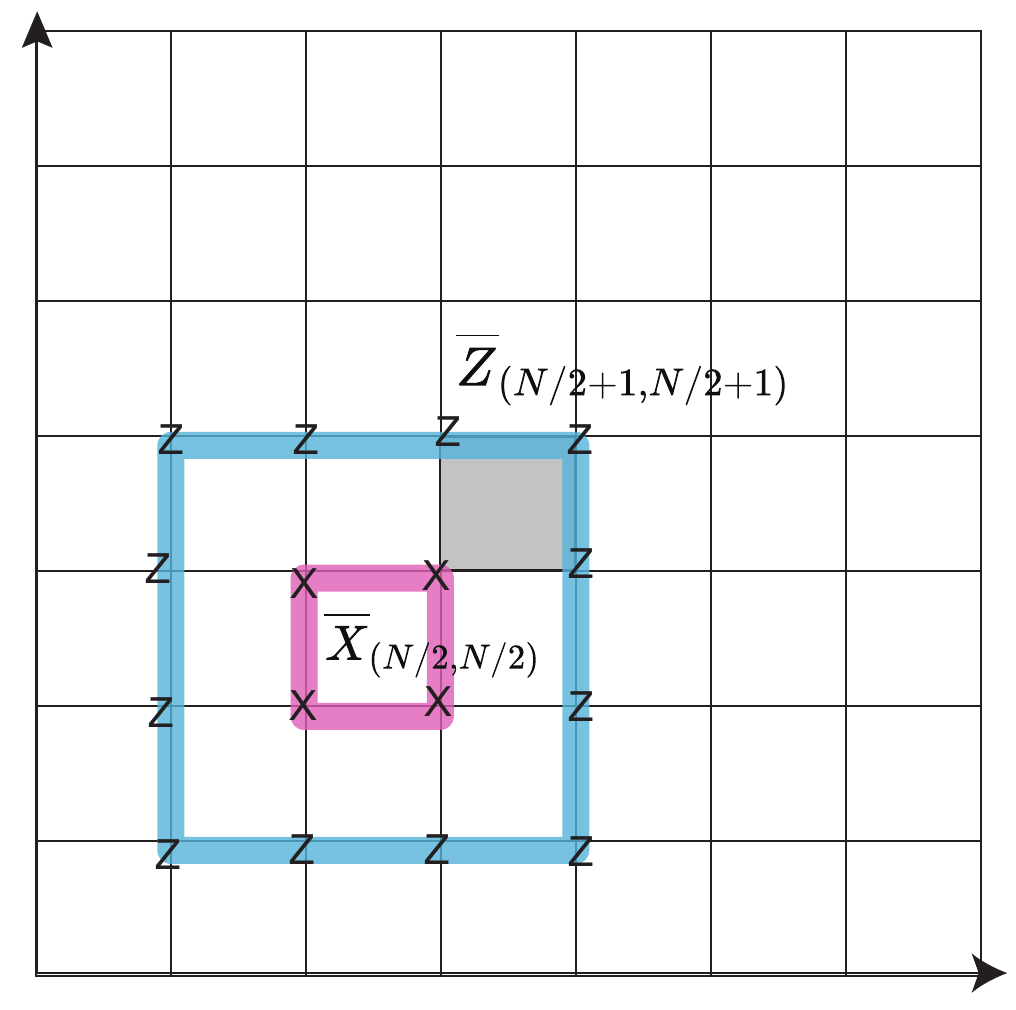}

}\subfloat[{Generators of controls ensemble $\{Y_{(N/2+3,j)}|j\in[1,N]\}$$\centering$}]{\includegraphics[width=4.5cm]{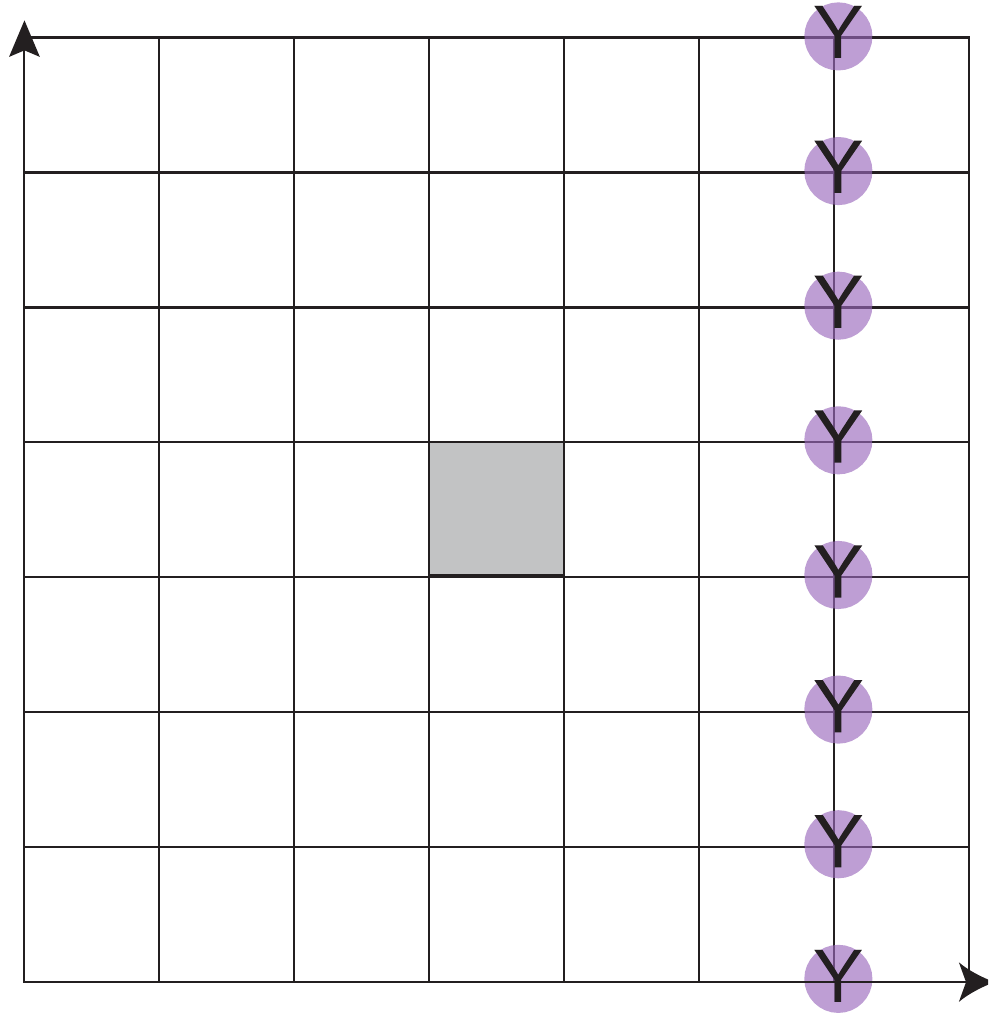}

}%
\end{minipage}

\caption{\label{SM:Sfig}Illustrations of TDC code. The gray square is $M_{0}$,
in the middle of the whole lattice.}
\end{figure*}

For clarity of description, we divide the $N\times N$ lattice into
three regions, 
\begin{equation}
L=G_{x}\cup G_{z}\cup M_{0},
\end{equation}
where $L$ denotes the set of all vertices in this lattice and $G_{x}$,
$G_{z}$, and $M_{0}$ are three disjoint subsets, as illustrated
in Fig. \ref{SM:fig:TDC-1}. Define
\begin{align*}
M_{0} & =\{(N/2,N/2),(N/2-1,N/2),(N/2,N/2-1),(N/2-1,N/2-1)\},\\
G_{x} & =\{(i,j)\in[1,N]^{2}|i+j\in2\mathbb{Z},(i,j)\notin M_{0}\},\\
G_{z} & =\{(i,j)\in[1,N]^{2}|i+j\notin2\mathbb{Z},(i,j)\notin M_{0}\}.
\end{align*}
The stabilizer group of the TDC code is 
\begin{equation}
\mathcal{S}=\left\langle S^{(x)}_{(i,j)},S^{(z)}_{(u,v)},\forall(i,j)\in G_{x},\forall(u,v)\in G_{z}\right\rangle ,
\end{equation}
where the stabilizer generators $S^{(x)}_{(i,j)}$ and $S^{(z)}_{(u,v)}$
consist of Pauli operator products resembling hollow crosses spanning
the entire lattice, plotted in Fig. \ref{SM:Sfig}, i.e.,
\begin{equation}
S^{(x)}_{(i,j)}\coloneqq\prod^{N}_{k=1,k\ne i}X_{(k,j)}\prod^{N}_{g=1,g\ne j}X_{(i,g)},\quad\forall(i,j)\in G_{x},
\end{equation}
\begin{equation}
S^{(z)}_{(u,v)}\coloneqq\prod^{N}_{k=1,k\ne u}Z_{(k,v)}\prod^{N}_{g=1,g\ne v}Z_{(u,g)},\quad\forall(u,v)\in G_{z},
\end{equation}
where $X_{(i,j)}$ represents the operator product that acts non-trivially
only on the $(i\text{-th},j\text{-th})$ qubit. And the logical operators
of this code are
\begin{equation}
\overline{X}_{(i,j)}\coloneqq\prod^{(1,1)}_{(u,v)=(0,0)}X_{(i+2u(i-\frac{N+1}{2}),(j+2v(j-\frac{N+1}{2}))},\quad\forall(i,j)\in M_{0},
\end{equation}
\begin{equation}
\overline{Z}_{(i,j)}\coloneqq\prod^{(3,3)}_{(u,v)=(0,0)}Z_{(i-2u(i-\frac{N+1}{2}),(j-2v(j-\frac{N+1}{2}))}\prod^{(1,1)}_{(0,0)}Z_{(i-2(u+1)(i-\frac{N+1}{2}),(j-2(v+1)(j-\frac{N+1}{2}))},\quad\forall(i,j)\in M_{0}.
\end{equation}

It can be readily verified that the operators $\{S^{(x)}_{(i,j)},S^{(z)}_{(u,v)}\}$
commute with all logical operators. Each logical operator $\overline{X}_{(i,j)}$
anticommutes with $\overline{Z}_{(i,j)},$ i.e., $\overline{X}_{(i,j)}\overline{Z}_{(i,j)}=-\overline{Z}_{(i,j)}\overline{X}_{(i,j)}$,
and commutes with all the other logical operators. Let $\mathcal{H}=(\mathbb{C}^{2})^{\otimes N^{2}}$
denote the joint Hilbert space of all $N^{2}$ qubits. Since the stabilizer
group $\mathcal{S}$ consists of a set of commuting observables, we
can use these observables to label subspaces of $\mathcal{H}$. In
particular we can label these subspaces by the $\pm1$ eigenvalues
of the $S^{(x)}_{(i,j)}$ and $S^{(z)}_{(u,v)}$ operators. Let us
denote these eigenvalues by $s^{(x)}_{(i,j)}$and $s^{(z)}_{(u,v)}$
respectively and the string with length $(N^{2}-4)/2$ of these \textpm 1
eigenvalues by $s^{(x)}$ and $s^{(z)}$. We can thus decompose $\mathcal{H}$
into subspaces as
\begin{equation}
\mathcal{H}=\bigoplus_{s^{(x)}_{(i,j)},s^{(z)}_{(u,v)}=\pm1}\mathcal{H}_{\left(s^{(x)}_{(i,j)},s^{(z)}_{(u,v)}\right)}=\bigoplus_{s^{(x)},s^{(z)}}\mathcal{H}_{(s^{(x)},s^{(z)})}.
\end{equation}
In total, there are $(N^{2}-4)$ stabilizer generators and $8$ logical
operators. Each of the $\mathcal{H}_{(s^{(x)},s^{(z)})}$ subspaces
has dimension $d=2^{N^{2}-(N^{2}-4)}=2^{4}$. Therefore, the code
encodes $4$ logical qubits into $N^{2}$ physical qubits. Now, consider
the stabilizer-sum Hamiltonian
\begin{equation}
H(\theta)=-\theta\left(\sum_{(i,j)\in G_{x}}S^{(x)}_{(i,j)}+\sum_{(u,v)\in G_{z}}S^{(z)}_{(u,v)}\right).
\end{equation}
For any fixed $i\in\{1,\cdots,N\}$, consider the set of symmetry
generators 
\[
\mathcal{A}_{i}:=\left\{ Y_{(i,j)}\mid j=1,\ldots,N\right\} .
\]
Each generator in $\mathcal{A}_{i}$ anticommutes with $O(N)$ stabilizer
generators, and collectively the generators in $\mathcal{A}_{i}$
cover all stabilizer generators. Thus, the total number $k$ of the generators of the symmetry transformation
is $N$, i.e., $k=N$. According to Eq. \eqref{SM:eq:sbound}, the bound
of QFI is
\begin{equation}
(\mathcal{F}_{\theta})_{\max}\ge\frac{4t^{2}}{N}(n-l)^{2}=O(n^{3/2}),
\end{equation}
which scales as the $3/2$-power of the number of physical qubits
$n$, exceeding the SQL!

\section{Symmetry optimization on five-qubit code}

According to the quantum Hamming bound, the five-qubit code is the
smallest QEC code which protects a logical qubit from any single-qubit
error. The choice of stabilizer code has some degree of freedom due
to the symmetry of one-qubit errors.

We consider the stabilizer group $\mathcal{S}=\left\langle g_{1},g_{2},g_{3},g_{4}\right\rangle $
generated by
\[
g_{1}=XZZXI,\quad g_{2}=IXZZX,\quad g_{3}=XIXZZ,\quad g_{4}=ZXIXZ.
\]
Two logical operators of the stabilizer code defined on $\mathcal{S}$
can be chosen as
\[
\overline{X}=XXXXX,\quad\overline{Z}=ZZZZZ.
\]
The logical space of the five-qubit code used in the main text is
spanned by
\begin{align}
\begin{aligned}\left|\bar{0}\right\rangle =\frac{1}{2^{4}}\prod^{4}_{i=1}(\mathbb{I}+g_{i})\left|00000\right\rangle = & \frac{1}{4}{[|00000\rangle+|10010\rangle+|01001\rangle+|10100\rangle+|01010\rangle-|11011\rangle-|00110\rangle-|11000\rangle}\\
 & -|11101\rangle-|00011\rangle-|11110\rangle-|01111\rangle-|10001\rangle-|01100\rangle-|10111\rangle+|00101\rangle]
\end{aligned}
\end{align}
and
\begin{figure}
\subfloat[\label{SM:fig:s_Optimization-of-QFIs}Optimization of QFI with respect
to $q$.\centering]{\includegraphics[width=9.5cm]{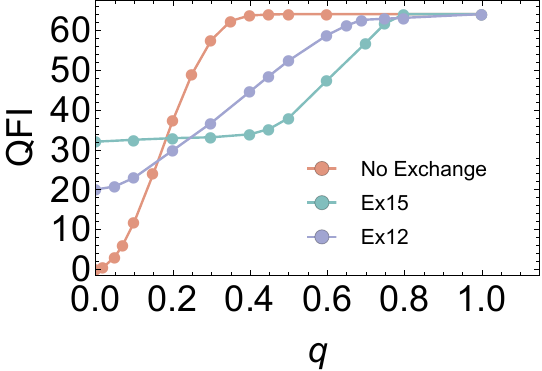}

}$\quad$$\quad$\subfloat[\label{SM:figS_:Optimal-control-sequence}Optimal control paths $\vec{r}_{1}(q)$
with and without qubit exchange\centering]{\includegraphics[width=7.4cm]{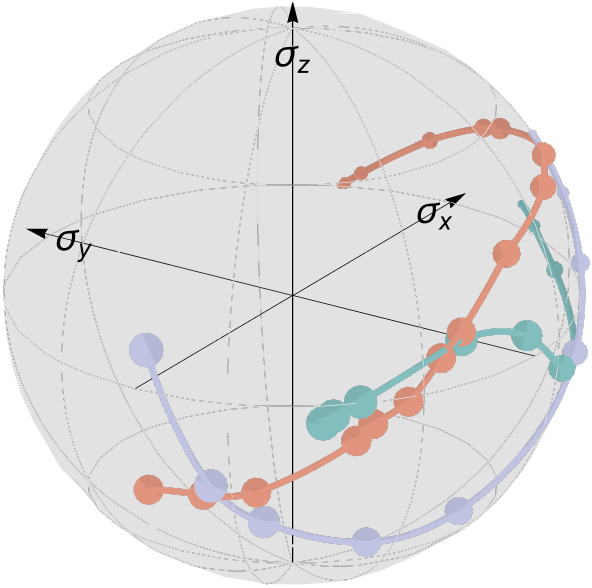}

}\caption{Symmetry optimization of five-qubit code for a stabilizer-sum Hamiltonian.
The symmetry transformation is $U(q)=\bigotimes_{k}e^{\text{i}\vec{r}_{k}(q)\cdot\vec{\sigma}}$,
where $\vec{r}_{1}(q)$ corresponds to the first qubit. (a) The efficacy
of optimization through continuous symmetry can be significantly enhanced
by exchange operations. Relative to continuous symmetry, operations
involving exchanging, such as exchanging the first and fifth qubits
(labeled as \textquotedblleft$\text{Ex}15$\textquotedblright ) as
well as exchanging the first and second qubits (labeled as \textquotedblleft$\text{Ex}12$\textquotedblright ),
can significantly increase the QFI at the initial evolution stages.
(b) The control paths $\vec{r}_{1}(q)$ are plotted and share the
legend with (a). The radius of points on the control paths represents
the estimation error.}
\end{figure}
\begin{equation}
\begin{aligned}\left|\bar{1}\right\rangle =\overline{X}\left|\bar{0}\right\rangle = & \frac{1}{4}[|11111\rangle+|01101\rangle+|10110\rangle+|01011\rangle+|10101\rangle-|00100\rangle-|11001\rangle-|00111\rangle\\
 & -|00010\rangle-|11100\rangle-|00001\rangle-|10000\rangle-|01110\rangle-|10011\rangle-|01000\rangle+|11010\rangle].
\end{aligned}
\end{equation}

Consider a stabilizer-sum Hamiltonian with an unknown prefactor $\theta$
to estimate,
\begin{equation}
H(\theta)=\theta H_{0},\;H_{0}=-\sum^{4}_{i=1}g_{i},
\end{equation}
which can be verified to meet the second-order HSO condition in Sec.
\ref{SM:subsec:Second-order-HSO-criterion} and therefore allow effective
symmetry optimization. Based on the symmetry-optimization framework
for QEC codes introduced in the previous sections, we use the following
symmetry transformations to optimize the QFI for the five-qubit code,
i.e.,
\begin{equation}
U(0,q_{t})=\bigotimes^{5}_{j=1}\mathcal{T\text{exp}}\left[-\text{i}\int^{q_{t}}_{0}A_{j}(q)dq\right],
\end{equation}
where $A_{j}(q)=\vec{r}_{j}(q)\cdot\vec{\sigma}_{j}$ acts nontrivially
only on the $j$-th physical qubit. The QFI becomes
\begin{equation}
\mathcal{F}_{\theta}=4t^{2}{\rm Var}_{U(0,q_{t})|\psi\rangle}(H_{0})=4t^{2}{\rm Var}_{|\psi\rangle}\left[U^{\dagger}(0,q_{t})H_{0}U(0,q_{t})\right],\label{SM:eq:optimization of QFI}
\end{equation}
with an initial state $|\psi\rangle=\alpha\left|\bar{0}\right\rangle +\beta\left|\bar{1}\right\rangle $.
The QFI reaches the maximum when $\vec{r}_{j}(q)$ takes specific
trajectories.

To achieve efficient numerical optimization, we discretize the evolution
of the system by the Trotter-Suzuki decomposition \cite{SM-Suzuki1976,SM-Trotter},
dividing the interval $I=[0,q_{t}]$ into $N$ segments, each spanning
$\Delta q=I/N$.  Assuming $N$ is sufficiently large so that $\Delta q$
is sufficiently small, $U(0,q_{t})$ can be approximated as
\begin{equation}
U(0,q_{t})=\bigotimes^{5}_{j=1}\prod^{N}_{k=1}\text{exp}\left[-\text{i}\vec{r}_{j}(k\Delta q)\cdot\vec{\sigma}_{j}\Delta q\right],
\end{equation}
which can be computed to arbitrary accuracy by standard numerical
optimization algorithms. By this approach, the continuous symmetry
optimization problem is transformed into a finite-dimensional optimization
problem, replacing the continuous path ordering by an ordered finite
product and enabling the use of existing quantum-control optimization
tools for simulations of quantum evolution, e.g., chopped random basis
(CRAB) \cite{SM-caneva2011chopped} and gradient ascent pulse engineering
(GRAPE) algorithm \cite{SM-khaneja2005optimal,SM-PhysRevA.103.023107,SM-PhysRevA.97.042122},
which we apply to optimize the symmetry transformation trajectories
in the main text.

\begingroup
\makeatletter
\let\NAT@bibsetnum\NATx@bibsetnum
\makeatother
\begin{NAT@thebibliography}{15}%
\makeatletter
\providecommand \@ifxundefined [1]{%
 \@ifx{#1\undefined}
}%
\providecommand \@ifnum [1]{%
 \ifnum #1\expandafter \@firstoftwo
 \else \expandafter \@secondoftwo
 \fi
}%
\providecommand \@ifx [1]{%
 \ifx #1\expandafter \@firstoftwo
 \else \expandafter \@secondoftwo
 \fi
}%
\providecommand \natexlab [1]{#1}%
\providecommand \enquote  [1]{``#1''}%
\providecommand \bibnamefont  [1]{#1}%
\providecommand \bibfnamefont [1]{#1}%
\providecommand \citenamefont [1]{#1}%
\providecommand \href@noop [0]{\@secondoftwo}%
\providecommand \href [0]{\begingroup \@sanitize@url \@href}%
\providecommand \@href[1]{\@@startlink{#1}\@@href}%
\providecommand \@@href[1]{\endgroup#1\@@endlink}%
\providecommand \@sanitize@url [0]{\catcode `\\12\catcode `\$12\catcode
  `\&12\catcode `\#12\catcode `\^12\catcode `\_12\catcode `\%12\relax}%
\providecommand \@@startlink[1]{}%
\providecommand \@@endlink[0]{}%
\providecommand \url  [0]{\begingroup\@sanitize@url \@url }%
\providecommand \@url [1]{\endgroup\@href {#1}{\urlprefix }}%
\providecommand \urlprefix  [0]{URL }%
\providecommand \Eprint [0]{\href }%
\providecommand \doibase [0]{https://doi.org/}%
\providecommand \selectlanguage [0]{\@gobble}%
\providecommand \bibinfo  [0]{\@secondoftwo}%
\providecommand \bibfield  [0]{\@secondoftwo}%
\providecommand \translation [1]{[#1]}%
\providecommand \BibitemOpen [0]{}%
\providecommand \bibitemStop [0]{}%
\providecommand \bibitemNoStop [0]{.\EOS\space}%
\providecommand \EOS [0]{\spacefactor3000\relax}%
\providecommand \BibitemShut  [1]{\csname bibitem#1\endcsname}%
\let\auto@bib@innerbib\@empty
%</preamble>
\bibitem [{\citenamefont {Fowler}\ \emph {et~al.}(2012)\citenamefont {Fowler},
  \citenamefont {Mariantoni}, \citenamefont {Martinis},\ and\ \citenamefont
  {Cleland}}]{SM-PhysRevA.86.032324}%
  \BibitemOpen
  \bibfield  {author} {\bibinfo {author} {\bibfnamefont {A.~G.}\ \bibnamefont
  {Fowler}}, \bibinfo {author} {\bibfnamefont {M.}~\bibnamefont {Mariantoni}},
  \bibinfo {author} {\bibfnamefont {J.~M.}\ \bibnamefont {Martinis}},\ and\
  \bibinfo {author} {\bibfnamefont {A.~N.}\ \bibnamefont {Cleland}},\ }\href
  {https://doi.org/10.1103/PhysRevA.86.032324} {\bibfield  {journal} {\bibinfo
  {journal} {Phys. Rev. A}\ }\textbf {\bibinfo {volume} {86}},\ \bibinfo
  {pages} {032324} (\bibinfo {year} {2012})}\BibitemShut {NoStop}%
\bibitem [{\citenamefont {Horsman}\ \emph {et~al.}(2012)\citenamefont
  {Horsman}, \citenamefont {Fowler}, \citenamefont {Devitt},\ and\
  \citenamefont {Van~Meter}}]{SM-Horsman2011SurfaceCQ}%
  \BibitemOpen
  \bibfield  {author} {\bibinfo {author} {\bibfnamefont {C.}~\bibnamefont
  {Horsman}}, \bibinfo {author} {\bibfnamefont {A.~G.}\ \bibnamefont {Fowler}},
  \bibinfo {author} {\bibfnamefont {S.}~\bibnamefont {Devitt}},\ and\ \bibinfo
  {author} {\bibfnamefont {R.}~\bibnamefont {Van~Meter}},\ }\href
  {https://doi.org/10.1088/1367-2630/14/12/123011} {\bibfield  {journal}
  {\bibinfo  {journal} {New J. Phys.}\ }\textbf {\bibinfo {volume} {14}},\
  \bibinfo {pages} {123011} (\bibinfo {year} {2012})}\BibitemShut {NoStop}%
\bibitem [{\citenamefont {Cleland}(2022)}]{SM-Cleland2022AnIT}%
  \BibitemOpen
  \bibfield  {author} {\bibinfo {author} {\bibfnamefont {A.~N.}\ \bibnamefont
  {Cleland}},\ }\href@noop {} {\bibfield  {journal} {\bibinfo  {journal}
  {SciPost Phys. Lect. Notes}\ }\textbf {\bibinfo {volume} {49}} (\bibinfo
  {year} {2022})}\BibitemShut {NoStop}%
\bibitem [{\citenamefont {Krinner}\ \emph {et~al.}(2022)\citenamefont
  {Krinner}, \citenamefont {Lacroix}, \citenamefont {Remm}, \citenamefont
  {Di~Paolo}, \citenamefont {Genois}, \citenamefont {Leroux}, \citenamefont
  {Hellings}, \citenamefont {Lazar}, \citenamefont {Swiadek}, \citenamefont
  {Herrmann}, \citenamefont {Norris}, \citenamefont {Andersen}, \citenamefont
  {M{\"u}ller}, \citenamefont {Blais}, \citenamefont {Eichler},\ and\
  \citenamefont {Wallraff}}]{SM-Krinner2022}%
  \BibitemOpen
  \bibfield  {author} {\bibinfo {author} {\bibfnamefont {S.}~\bibnamefont
  {Krinner}}, \bibinfo {author} {\bibfnamefont {N.}~\bibnamefont {Lacroix}},
  \bibinfo {author} {\bibfnamefont {A.}~\bibnamefont {Remm}}, \bibinfo {author}
  {\bibfnamefont {A.}~\bibnamefont {Di~Paolo}}, \bibinfo {author}
  {\bibfnamefont {E.}~\bibnamefont {Genois}}, \bibinfo {author} {\bibfnamefont
  {C.}~\bibnamefont {Leroux}}, \bibinfo {author} {\bibfnamefont
  {C.}~\bibnamefont {Hellings}}, \bibinfo {author} {\bibfnamefont
  {S.}~\bibnamefont {Lazar}}, \bibinfo {author} {\bibfnamefont
  {F.}~\bibnamefont {Swiadek}}, \bibinfo {author} {\bibfnamefont
  {J.}~\bibnamefont {Herrmann}}, \bibinfo {author} {\bibfnamefont {G.~J.}\
  \bibnamefont {Norris}}, \bibinfo {author} {\bibfnamefont {C.~K.}\
  \bibnamefont {Andersen}}, \bibinfo {author} {\bibfnamefont {M.}~\bibnamefont
  {M{\"u}ller}}, \bibinfo {author} {\bibfnamefont {A.}~\bibnamefont {Blais}},
  \bibinfo {author} {\bibfnamefont {C.}~\bibnamefont {Eichler}},\ and\ \bibinfo
  {author} {\bibfnamefont {A.}~\bibnamefont {Wallraff}},\ }\href
  {https://doi.org/10.1038/s41586-022-04566-8} {\bibfield  {journal} {\bibinfo
  {journal} {Nature}\ }\textbf {\bibinfo {volume} {605}},\ \bibinfo {pages}
  {669} (\bibinfo {year} {2022})}\BibitemShut {NoStop}%
\bibitem [{\citenamefont {Acharya}\ \emph {et~al.}(2025)\citenamefont {Acharya}
  \emph {et~al.}}]{SM-Acharya2025}%
  \BibitemOpen
  \bibfield  {author} {\bibinfo {author} {\bibfnamefont {R.}~\bibnamefont
  {Acharya}} \emph {et~al.},\ }\href
  {https://doi.org/10.1038/s41586-024-08449-y} {\bibfield  {journal} {\bibinfo
  {journal} {Nature}\ }\textbf {\bibinfo {volume} {638}},\ \bibinfo {pages}
  {920} (\bibinfo {year} {2025})}\BibitemShut {NoStop}%
\bibitem [{\citenamefont {Zhao}\ \emph {et~al.}(2022)\citenamefont {Zhao},
  \citenamefont {Ye}, \citenamefont {Huang}, \citenamefont {Zhang},
  \citenamefont {Wu}, \citenamefont {Guan}, \citenamefont {Zhu}, \citenamefont
  {Wei}, \citenamefont {He}, \citenamefont {Cao}, \citenamefont {Chen},
  \citenamefont {Chung}, \citenamefont {Deng}, \citenamefont {Fan},
  \citenamefont {Gong}, \citenamefont {Guo}, \citenamefont {Guo}, \citenamefont
  {Han}, \citenamefont {Li}, \citenamefont {Li}, \citenamefont {Li},
  \citenamefont {Liang}, \citenamefont {Lin}, \citenamefont {Qian},
  \citenamefont {Rong}, \citenamefont {Su}, \citenamefont {Sun}, \citenamefont
  {Wang}, \citenamefont {Wu}, \citenamefont {Xu}, \citenamefont {Ying},
  \citenamefont {Yu}, \citenamefont {Zha}, \citenamefont {Zhang}, \citenamefont
  {Huo}, \citenamefont {Lu}, \citenamefont {Peng}, \citenamefont {Zhu},\ and\
  \citenamefont {Pan}}]{SM-zhaoRealizationErrorCorrectingSurface2022a}%
  \BibitemOpen
  \bibfield  {author} {\bibinfo {author} {\bibfnamefont {Y.}~\bibnamefont
  {Zhao}}, \bibinfo {author} {\bibfnamefont {Y.}~\bibnamefont {Ye}}, \bibinfo
  {author} {\bibfnamefont {H.-L.}\ \bibnamefont {Huang}}, \bibinfo {author}
  {\bibfnamefont {Y.}~\bibnamefont {Zhang}}, \bibinfo {author} {\bibfnamefont
  {D.}~\bibnamefont {Wu}}, \bibinfo {author} {\bibfnamefont {H.}~\bibnamefont
  {Guan}}, \bibinfo {author} {\bibfnamefont {Q.}~\bibnamefont {Zhu}}, \bibinfo
  {author} {\bibfnamefont {Z.}~\bibnamefont {Wei}}, \bibinfo {author}
  {\bibfnamefont {T.}~\bibnamefont {He}}, \bibinfo {author} {\bibfnamefont
  {S.}~\bibnamefont {Cao}}, \bibinfo {author} {\bibfnamefont {F.}~\bibnamefont
  {Chen}}, \bibinfo {author} {\bibfnamefont {T.-H.}\ \bibnamefont {Chung}},
  \bibinfo {author} {\bibfnamefont {H.}~\bibnamefont {Deng}}, \bibinfo {author}
  {\bibfnamefont {D.}~\bibnamefont {Fan}}, \bibinfo {author} {\bibfnamefont
  {M.}~\bibnamefont {Gong}}, \bibinfo {author} {\bibfnamefont {C.}~\bibnamefont
  {Guo}}, \bibinfo {author} {\bibfnamefont {S.}~\bibnamefont {Guo}}, \bibinfo
  {author} {\bibfnamefont {L.}~\bibnamefont {Han}}, \bibinfo {author}
  {\bibfnamefont {N.}~\bibnamefont {Li}}, \bibinfo {author} {\bibfnamefont
  {S.}~\bibnamefont {Li}}, \bibinfo {author} {\bibfnamefont {Y.}~\bibnamefont
  {Li}}, \bibinfo {author} {\bibfnamefont {F.}~\bibnamefont {Liang}}, \bibinfo
  {author} {\bibfnamefont {J.}~\bibnamefont {Lin}}, \bibinfo {author}
  {\bibfnamefont {H.}~\bibnamefont {Qian}}, \bibinfo {author} {\bibfnamefont
  {H.}~\bibnamefont {Rong}}, \bibinfo {author} {\bibfnamefont {H.}~\bibnamefont
  {Su}}, \bibinfo {author} {\bibfnamefont {L.}~\bibnamefont {Sun}}, \bibinfo
  {author} {\bibfnamefont {S.}~\bibnamefont {Wang}}, \bibinfo {author}
  {\bibfnamefont {Y.}~\bibnamefont {Wu}}, \bibinfo {author} {\bibfnamefont
  {Y.}~\bibnamefont {Xu}}, \bibinfo {author} {\bibfnamefont {C.}~\bibnamefont
  {Ying}}, \bibinfo {author} {\bibfnamefont {J.}~\bibnamefont {Yu}}, \bibinfo
  {author} {\bibfnamefont {C.}~\bibnamefont {Zha}}, \bibinfo {author}
  {\bibfnamefont {K.}~\bibnamefont {Zhang}}, \bibinfo {author} {\bibfnamefont
  {Y.-H.}\ \bibnamefont {Huo}}, \bibinfo {author} {\bibfnamefont {C.-Y.}\
  \bibnamefont {Lu}}, \bibinfo {author} {\bibfnamefont {C.-Z.}\ \bibnamefont
  {Peng}}, \bibinfo {author} {\bibfnamefont {X.}~\bibnamefont {Zhu}},\ and\
  \bibinfo {author} {\bibfnamefont {J.-W.}\ \bibnamefont {Pan}},\ }\href
  {https://doi.org/10.1103/PhysRevLett.129.030501} {\bibfield  {journal}
  {\bibinfo  {journal} {Phys. Rev. Lett.}\ }\textbf {\bibinfo {volume} {129}},\
  \bibinfo {pages} {030501} (\bibinfo {year} {2022})}\BibitemShut {NoStop}%
\bibitem [{\citenamefont {Paik}\ \emph {et~al.}(2011)\citenamefont {Paik},
  \citenamefont {Schuster}, \citenamefont {Bishop}, \citenamefont {Kirchmair},
  \citenamefont {Catelani}, \citenamefont {Sears}, \citenamefont {Johnson},
  \citenamefont {Reagor}, \citenamefont {Frunzio}, \citenamefont {Glazman},
  \citenamefont {Girvin}, \citenamefont {Devoret},\ and\ \citenamefont
  {Schoelkopf}}]{SM-PhysRevLett.107.240501}%
  \BibitemOpen
  \bibfield  {author} {\bibinfo {author} {\bibfnamefont {H.}~\bibnamefont
  {Paik}}, \bibinfo {author} {\bibfnamefont {D.~I.}\ \bibnamefont {Schuster}},
  \bibinfo {author} {\bibfnamefont {L.~S.}\ \bibnamefont {Bishop}}, \bibinfo
  {author} {\bibfnamefont {G.}~\bibnamefont {Kirchmair}}, \bibinfo {author}
  {\bibfnamefont {G.}~\bibnamefont {Catelani}}, \bibinfo {author}
  {\bibfnamefont {A.~P.}\ \bibnamefont {Sears}}, \bibinfo {author}
  {\bibfnamefont {B.~R.}\ \bibnamefont {Johnson}}, \bibinfo {author}
  {\bibfnamefont {M.~J.}\ \bibnamefont {Reagor}}, \bibinfo {author}
  {\bibfnamefont {L.}~\bibnamefont {Frunzio}}, \bibinfo {author} {\bibfnamefont
  {L.~I.}\ \bibnamefont {Glazman}}, \bibinfo {author} {\bibfnamefont {S.~M.}\
  \bibnamefont {Girvin}}, \bibinfo {author} {\bibfnamefont {M.~H.}\
  \bibnamefont {Devoret}},\ and\ \bibinfo {author} {\bibfnamefont {R.~J.}\
  \bibnamefont {Schoelkopf}},\ }\href
  {https://doi.org/10.1103/PhysRevLett.107.240501} {\bibfield  {journal}
  {\bibinfo  {journal} {Phys. Rev. Lett.}\ }\textbf {\bibinfo {volume} {107}},\
  \bibinfo {pages} {240501} (\bibinfo {year} {2011})}\BibitemShut {NoStop}%
\bibitem [{\citenamefont {Mariantoni}\ \emph {et~al.}(2011)\citenamefont
  {Mariantoni}, \citenamefont {Wang}, \citenamefont {Yamamoto}, \citenamefont
  {Neeley}, \citenamefont {Bialczak}, \citenamefont {Chen}, \citenamefont
  {Lenander}, \citenamefont {Lucero}, \citenamefont {O'Connell}, \citenamefont
  {Sank}, \citenamefont {Weides}, \citenamefont {Wenner}, \citenamefont {Yin},
  \citenamefont {Zhao}, \citenamefont {Korotkov}, \citenamefont {Cleland},\
  and\ \citenamefont {Martinis}}]{SM-Mariantoni2011ImplementingTQ}%
  \BibitemOpen
  \bibfield  {author} {\bibinfo {author} {\bibfnamefont {M.}~\bibnamefont
  {Mariantoni}}, \bibinfo {author} {\bibfnamefont {H.}~\bibnamefont {Wang}},
  \bibinfo {author} {\bibfnamefont {T.}~\bibnamefont {Yamamoto}}, \bibinfo
  {author} {\bibfnamefont {M.}~\bibnamefont {Neeley}}, \bibinfo {author}
  {\bibfnamefont {R.~C.}\ \bibnamefont {Bialczak}}, \bibinfo {author}
  {\bibfnamefont {Y.}~\bibnamefont {Chen}}, \bibinfo {author} {\bibfnamefont
  {M.}~\bibnamefont {Lenander}}, \bibinfo {author} {\bibfnamefont
  {E.}~\bibnamefont {Lucero}}, \bibinfo {author} {\bibfnamefont {A.~D.}\
  \bibnamefont {O'Connell}}, \bibinfo {author} {\bibfnamefont {D.~T.}\
  \bibnamefont {Sank}}, \bibinfo {author} {\bibfnamefont {M.~P.}\ \bibnamefont
  {Weides}}, \bibinfo {author} {\bibfnamefont {J.}~\bibnamefont {Wenner}},
  \bibinfo {author} {\bibfnamefont {Y.}~\bibnamefont {Yin}}, \bibinfo {author}
  {\bibfnamefont {J.}~\bibnamefont {Zhao}}, \bibinfo {author} {\bibfnamefont
  {A.~N.}\ \bibnamefont {Korotkov}}, \bibinfo {author} {\bibfnamefont {A.~N.}\
  \bibnamefont {Cleland}},\ and\ \bibinfo {author} {\bibfnamefont {J.~M.}\
  \bibnamefont {Martinis}},\ }\href
  {https://api.semanticscholar.org/CorpusID:11483576} {\bibfield  {journal}
  {\bibinfo  {journal} {Science}\ }\textbf {\bibinfo {volume} {334}},\ \bibinfo
  {pages} {61 } (\bibinfo {year} {2011})}\BibitemShut {NoStop}%
\bibitem [{\citenamefont {Bluvstein}\ \emph {et~al.}(2022)\citenamefont
  {Bluvstein}, \citenamefont {Levine}, \citenamefont {Semeghini}, \citenamefont
  {Wang}, \citenamefont {Ebadi}, \citenamefont {Kalinowski}, \citenamefont
  {Keesling}, \citenamefont {Maskara}, \citenamefont {Pichler}, \citenamefont
  {Greiner}, \citenamefont {Vuleti{\'{c}}},\ and\ \citenamefont
  {Lukin}}]{SM-Bluvstein2022}%
  \BibitemOpen
  \bibfield  {author} {\bibinfo {author} {\bibfnamefont {D.}~\bibnamefont
  {Bluvstein}}, \bibinfo {author} {\bibfnamefont {H.}~\bibnamefont {Levine}},
  \bibinfo {author} {\bibfnamefont {G.}~\bibnamefont {Semeghini}}, \bibinfo
  {author} {\bibfnamefont {T.~T.}\ \bibnamefont {Wang}}, \bibinfo {author}
  {\bibfnamefont {S.}~\bibnamefont {Ebadi}}, \bibinfo {author} {\bibfnamefont
  {M.}~\bibnamefont {Kalinowski}}, \bibinfo {author} {\bibfnamefont
  {A.}~\bibnamefont {Keesling}}, \bibinfo {author} {\bibfnamefont
  {N.}~\bibnamefont {Maskara}}, \bibinfo {author} {\bibfnamefont
  {H.}~\bibnamefont {Pichler}}, \bibinfo {author} {\bibfnamefont
  {M.}~\bibnamefont {Greiner}}, \bibinfo {author} {\bibfnamefont
  {V.}~\bibnamefont {Vuleti{\'{c}}}},\ and\ \bibinfo {author} {\bibfnamefont
  {M.~D.}\ \bibnamefont {Lukin}},\ }\href
  {https://doi.org/10.1038/s41586-022-04592-6} {\bibfield  {journal} {\bibinfo
  {journal} {Nature}\ }\textbf {\bibinfo {volume} {604}},\ \bibinfo {pages}
  {451} (\bibinfo {year} {2022})}\BibitemShut {NoStop}%
\bibitem [{\citenamefont {Suzuki}(1976)}]{SM-Suzuki1976}%
  \BibitemOpen
  \bibfield  {author} {\bibinfo {author} {\bibfnamefont {M.}~\bibnamefont
  {Suzuki}},\ }\href {https://doi.org/10.1007/BF01609348} {\bibfield  {journal}
  {\bibinfo  {journal} {Commun. Math. Phys.}\ }\textbf {\bibinfo {volume}
  {51}},\ \bibinfo {pages} {183} (\bibinfo {year} {1976})}\BibitemShut
  {NoStop}%
\bibitem [{\citenamefont {Trotter}(1959)}]{SM-Trotter}%
  \BibitemOpen
  \bibfield  {author} {\bibinfo {author} {\bibfnamefont {H.~F.}\ \bibnamefont
  {Trotter}},\ }\href {http://www.jstor.org/stable/2033649} {\bibfield
  {journal} {\bibinfo  {journal} {Proc. Am. Math. Soc.}\ }\textbf {\bibinfo
  {volume} {10}},\ \bibinfo {pages} {545} (\bibinfo {year} {1959})}\BibitemShut
  {NoStop}%
\bibitem [{\citenamefont {Caneva}\ \emph {et~al.}(2011)\citenamefont {Caneva},
  \citenamefont {Calarco},\ and\ \citenamefont
  {Montangero}}]{SM-caneva2011chopped}%
  \BibitemOpen
  \bibfield  {author} {\bibinfo {author} {\bibfnamefont {T.}~\bibnamefont
  {Caneva}}, \bibinfo {author} {\bibfnamefont {T.}~\bibnamefont {Calarco}},\
  and\ \bibinfo {author} {\bibfnamefont {S.}~\bibnamefont {Montangero}},\
  }\href@noop {} {\bibfield  {journal} {\bibinfo  {journal} {Phys. Rev. A}\
  }\textbf {\bibinfo {volume} {84}},\ \bibinfo {pages} {022326} (\bibinfo
  {year} {2011})}\BibitemShut {NoStop}%
\bibitem [{\citenamefont {Khaneja}\ \emph {et~al.}(2005)\citenamefont
  {Khaneja}, \citenamefont {Reiss}, \citenamefont {Kehlet}, \citenamefont
  {Schulte-Herbr{\"u}ggen},\ and\ \citenamefont {Glaser}}]{SM-khaneja2005optimal}%
  \BibitemOpen
  \bibfield  {author} {\bibinfo {author} {\bibfnamefont {N.}~\bibnamefont
  {Khaneja}}, \bibinfo {author} {\bibfnamefont {T.}~\bibnamefont {Reiss}},
  \bibinfo {author} {\bibfnamefont {C.}~\bibnamefont {Kehlet}}, \bibinfo
  {author} {\bibfnamefont {T.}~\bibnamefont {Schulte-Herbr{\"u}ggen}},\ and\
  \bibinfo {author} {\bibfnamefont {S.~J.}\ \bibnamefont {Glaser}},\
  }\href@noop {} {\bibfield  {journal} {\bibinfo  {journal} {J. Magn. Reson.}\
  }\textbf {\bibinfo {volume} {172}},\ \bibinfo {pages} {296} (\bibinfo {year}
  {2005})}\BibitemShut {NoStop}%
\bibitem [{\citenamefont {Larocca}\ and\ \citenamefont
  {Wisniacki}(2021)}]{SM-PhysRevA.103.023107}%
  \BibitemOpen
  \bibfield  {author} {\bibinfo {author} {\bibfnamefont {M.}~\bibnamefont
  {Larocca}}\ and\ \bibinfo {author} {\bibfnamefont {D.}~\bibnamefont
  {Wisniacki}},\ }\href {https://doi.org/10.1103/PhysRevA.103.023107}
  {\bibfield  {journal} {\bibinfo  {journal} {Phys. Rev. A}\ }\textbf {\bibinfo
  {volume} {103}},\ \bibinfo {pages} {023107} (\bibinfo {year}
  {2021})}\BibitemShut {NoStop}%
\bibitem [{\citenamefont {Wu}\ \emph {et~al.}(2018)\citenamefont {Wu},
  \citenamefont {Chu}, \citenamefont {Owens},\ and\ \citenamefont
  {Rabitz}}]{SM-PhysRevA.97.042122}%
  \BibitemOpen
  \bibfield  {author} {\bibinfo {author} {\bibfnamefont {R.-B.}\ \bibnamefont
  {Wu}}, \bibinfo {author} {\bibfnamefont {B.}~\bibnamefont {Chu}}, \bibinfo
  {author} {\bibfnamefont {D.~H.}\ \bibnamefont {Owens}},\ and\ \bibinfo
  {author} {\bibfnamefont {H.}~\bibnamefont {Rabitz}},\ }\href
  {https://doi.org/10.1103/PhysRevA.97.042122} {\bibfield  {journal} {\bibinfo
  {journal} {Phys. Rev. A}\ }\textbf {\bibinfo {volume} {97}},\ \bibinfo
  {pages} {042122} (\bibinfo {year} {2018})}\BibitemShut {NoStop}%
\makeatother
\end{NAT@thebibliography}%
\endgroup
\makeatother

\end{document}